\documentclass[iop]{emulateapj}

\usepackage{graphicx}
\usepackage{lineno}                   
\usepackage{amsmath}                  
\usepackage{dcolumn}                  
\usepackage{bm}                       
\usepackage{xspace}                   
\usepackage{hyperref}                 

\begin{document}

\shorttitle{Cosmic-Ray Anisotropy in IceCube}
\shortauthors{M.~G.~Aartsen et al.} 

\title{Anisotropy in Cosmic-Ray Arrival Directions in the Southern Hemisphere Based on Six Years of Data from
the IceCube Detector}

\author{
IceCube Collaboration:
M.~G.~Aartsen\altaffilmark{1},
K.~Abraham\altaffilmark{2},
M.~Ackermann\altaffilmark{3},
J.~Adams\altaffilmark{4},
J.~A.~Aguilar\altaffilmark{5},
M.~Ahlers\altaffilmark{6},
M.~Ahrens\altaffilmark{7},
D.~Altmann\altaffilmark{8},
T.~Anderson\altaffilmark{9},
I.~Ansseau\altaffilmark{5}
% ...
}
\altaffiltext{1}{Department of Physics, University of Adelaide, Adelaide, 5005, Australia}
\altaffiltext{2}{Technische Universit\"at M\"unchen, D-85748 Garching, Germany}
\altaffiltext{3}{DESY, D-15735 Zeuthen, Germany}
\altaffiltext{4}{Dept.~of Physics and Astronomy, University of Canterbury, Private Bag 4800, Christchurch, New Zealand}
\altaffiltext{5}{Universit\'e Libre de Bruxelles, Science Faculty CP230, B-1050 Brussels, Belgium}
\altaffiltext{6}{Dept.~of Physics and Wisconsin IceCube Particle Astrophysics Center, University of Wisconsin, Madison, WI 53706, USA}
\altaffiltext{7}{Oskar Klein Centre and Dept.~of Physics, Stockholm University, SE-10691 Stockholm, Sweden}
\altaffiltext{8}{Erlangen Centre for Astroparticle Physics, Friedrich-Alexander-Universit\"at Erlangen-N\"urnberg, D-91058 Erlangen, Germany}
\altaffiltext{9}{Dept.~of Physics, Pennsylvania State University, University Park, PA 16802, USA}
% ...


\begin{abstract}
The IceCube Neutrino Observatory has accumulated a total of 318 billion cosmic-ray
induced muon events between May 2009 and May 2015.  This data set was used for a 
detailed analysis of the sidereal cosmic-ray arrival direction anisotropy in the TeV to PeV 
energy range.  The observed global sidereal anisotropy features large regions of relative 
excess and deficit, with amplitudes on the order of $10^{-3}$ up to about 100 TeV.  
A decomposition of the arrival direction distribution into spherical harmonics 
shows that most of the power is contained in the low-multipole ($\ell\leq 4$) 
moments.  However, higher multipole components are found to be statistically 
significant down to an angular scale of less than $10^{\circ}$, approaching the 
angular resolution of the detector.  Above 100 TeV, a change in the morphology of 
the arrival direction distribution is observed, and the anisotropy is characterized 
by a wide relative deficit whose amplitude increases with primary energy up to 
at least 5\,PeV, the highest energies currently accessible to IceCube.  No 
time dependence of the large- and small-scale structures is observed in the six-year 
period covered by this analysis.  The high-statistics data set reveals more details 
on the properties of the anisotropy and is potentially able to shed light on the various 
physical processes that are responsible for the complex angular structure and
energy evolution.
\end{abstract}
{\keywords{astroparticle physics --- cosmic rays}


\section{Introduction}
\label{sec:introduction}

In the last few decades, a number of experiments have provided long-term,
statistically significant evidence of a faint sidereal anisotropy in the cosmic-ray
arrival direction distribution across six orders of magnitude in energy, from tens of GeV
to tens of PeV. The small amplitude of the observed large-scale anisotropy, on the
order of $10^{-4}$--$10^{-3}$, alongside the energy-dependent morphology and angular
structure, hint at multiple phenomenological contributions to the observations.
Muon detectors in the Northern Hemisphere observed cosmic-ray anisotropy from
energies of several tens to hundreds of GeV, which is beyond the direct solar
wind influence~\citep{Nagashima:1998aug, Hall:1999apr, Matsushiro:2010apr}.
Various surface arrays and underground detectors reported observations in the
TeV energy range~\citep{Tibet:2005jun, Tibet:2006oct, SuperK:2007mar, 
Milagro:2008nov, Milagro:2009jun, MINOS:2011icrc, Oshima:2011, ARGO:2013oct, 
Bartoli:2015, HAWC:2014dec}, in some cases extending up to hundreds of 
TeV~\citep{Aglietta:2009feb, KASCADE:2014aug, Tibet:2015}.
Recently, the IceCube Neutrino Observatory reported the first observations of 
sidereal cosmic-ray anisotropy in the Southern Hemisphere, with unprecedented event 
statistics in the energy range between 10\,TeV and 1\,PeV~\citep{IceCube:2010aug,
IceCube:2011oct, IceCube:2012feb, IceCube:2013mar}.

All the observations show a morphologically consistent large-scale anisotropy
structure across the sky in celestial coordinates.  In the energy interval from
60\,GeV to 100\,TeV, the cosmic-ray arrival direction distribution shows a wide
relative excess in the 30$^{\circ}$--120$^{\circ}$ range in right ascension, and
a deficit in the 150$^{\circ}$--250$^{\circ}$ range.  The degree of this
directional asymmetry is found to increase with energy up to about 10\,TeV, and
to decrease at higher energies up to about 100\,TeV.  In the 100--300\,TeV energy
range, a major change in the anisotropy morphology is observed by both the EAS-TOP 
array in the Northern Hemisphere~\citep{Aglietta:2009feb} and by IceCube in the 
Southern Hemisphere~\citep{IceCube:2012feb}.  IceCube data show that at an energy 
in excess of a few hundred TeV, the cosmic-ray arrival distribution seems 
mostly characterized by a relative deficit at a right ascension of 
60$^{\circ}$--120$^{\circ}$~\citep{IceCube:2012feb}, with amplitude increasing 
with energy.

It is evident from the observations that the anisotropy cannot be described with 
a simple dipole, although this is typically used to estimate the amplitude. 
Instead, a quantitative description of the anisotropy requires a characterization 
of the distribution over a wide range of angular scales.  In particular, the 
arrival direction distribution can be described as a superposition of spherical 
harmonic contributions, where most of the power is in the low-multipole 
($\ell \leq 4$) terms.  A fit using only the low-multipole terms, however, describes 
the data poorly, indicating that the higher-multipole terms must be accounted for 
as well.  In fact, statistically significant smaller angular scale features, with 
amplitudes on the order of $10^{-5}$--$10^{-4}$, have been observed in the TeV 
energy range by several experiments~\citep{Tibet:2007aug, Milagro:2008nov, 
IceCube:2011oct, ARGO:2013oct, HAWC:2014dec}.  More detailed studies with detectors 
in the Northern Hemisphere have shown that the energy spectrum of the cosmic-ray 
flux in the most dominant excess region is harder than that of the isotropic 
cosmic-ray flux~\citep{Milagro:2008nov, ARGO:2013oct, HAWC:2014dec}.  Although further 
confirmation is needed, this result might indicate that whatever produces the localized 
region of excess is responsible for the spectral deviation as well.

The evolution of cosmic-ray anisotropy in energy has also been observed by
experiments sensitive to ultra-high-energy particles.  The Pierre Auger
Observatory, in a search for a dipole and quadrupole component of the
cosmic-ray arrival direction distribution, found that a shift in the phase of the 
anisotropy occurs at about 1\,EeV as well~\citep{Auger:2011mar, Auger:2012dec, 
Auger:2013jan, Aab:2015bza}.  Below 1\,EeV, the dipole phase is consistent
with the phase observed by IceCube at PeV energies.  Around 4\,EeV, the
phase changes and the relative excess moves towards the range in right 
ascension that includes the Galactic anticenter direction.  This may be an 
indication that a new population of extragalactic cosmic rays begins to become 
dominant.  At energies in excess of 1\,EeV, cosmic rays are less bent by Galactic 
and intergalactic magnetic fields, making it possible to transition into the 
regime of cosmic-ray astronomy~\citep{Abbasi:2014lda} if the composition is light.

This paper reports new results on the energy and time dependence of the cosmic-ray 
anisotropy as observed by IceCube.  It is based on 318 billion cosmic-ray 
events recorded between May 2009 and May 2015.  
The large size of the data set allows for a detailed study of the energy
dependence of the anisotropy in the TeV to PeV energy range.  At PeV energies,
additional data from the IceTop air-shower array is used to provide an independent
analysis.  We also include a study of the time dependence of the anisotropy over 
the six-year period of data-taking used in this analysis.

The paper is organized as follows.  In Section\,\ref{sec:icecube}, we describe the 
IceCube detector and summarize basic characteristics of the data set used in 
this analysis.  The analysis techniques, including the energy estimation for the 
cosmic-ray primaries, are described in Section\,\ref{sec:analysis}.  Results on large
and small-scale anisotropy, including a study of the anisotropy in several energy 
bands from 13\,TeV to 5.3\,PeV and a study of the stability of the anisotropy,
are reported in Section~\ref{sec:results}.  Several systematic checks are described 
in Section\,\ref{sec:systematics}.  A discussion of the results 
(Section\,\ref{sec:discussion}) concludes the paper.  Many of the techniques used 
in this analysis are described in detail in~\citet{IceCube:2010aug, IceCube:2011oct, 
IceCube:2012feb} and~\citet{IceCube:2013mar}.

\section{IceCube}
\label{sec:icecube}

\begin{table*}[ht]
  \centering
  \begin{tabular}{crr}
    \hline
    Configuration & Livetime (days) & Number of Events \\
    \hline
    IC59       &   339.38 (91.7\%) & $3.579 \times 10^{10}$ \\
    IC79       &   315.76 (92.6\%) & $4.131 \times 10^{10}$ \\
    IC86-I     &   343.04 (93.0\%) & $5.906 \times 10^{10}$ \\
    IC86-II    &   331.92 (94.0\%) & $5.630 \times 10^{10}$ \\
    IC86-III   &   362.20 (97.9\%) & $6.214 \times 10^{10}$ \\
    IC86-IV    &   369.76 (97.8\%) & $6.327 \times 10^{10}$ \\
		\hline
    Total      &  2062.06 (94.5\%) & $3.179 \times 10^{11}$ \\
    \hline
  \end{tabular}
  \begin{tabular}{crr}
    \hline
    Configuration  &  Livetime (days)  &  Number of Events \\
    \hline
    IT59       &   338.25 (91.4\%) & $2.887 \times 10^{7}$ \\
    IT73       &   312.66 (91.7\%) & $3.690 \times 10^{7}$ \\
    IT81       &   343.04 (93.0\%) & $3.800 \times 10^{7}$ \\
    IT81-II    &   332.26 (94.1\%) & $3.713 \times 10^{7}$ \\
    IT81-III   &   361.20 (97.6\%) & $3.101 \times 10^{7}$ \\
    IT81-IV    &   362.61 (95.9\%) & $2.810 \times 10^{7}$ \\
    \hline
    Total      &  2050.01 (94.0\%) & $1.719 \times 10^{8}$ \\
    \hline
  \end{tabular}
  \caption{Detector configurations and their respective number of events for
all years used in this analysis. {\it IC} indicates IceCube, {\it IT} IceTop, 
and the number that follows indicates the number of strings or stations 
participating in data acquisition.}
  \label{tab:data}
\end{table*}

The IceCube Neutrino Observatory~\citep{IceCube:2006oct}, located at the 
geographic South Pole, comprises a neutrino detector in the deep ice 
(hereafter labeled IceCube) and a surface air-shower array (labeled IceTop).
Completed in 2010 after seven years of construction, IceCube consists of 86
vertical strings containing a total of 5,160 optical sensors, called digital
optical modules (DOMs), frozen in the ice at depths from 1.5 to 2.5\,km 
below the surface of the ice.  A DOM consists of a pressure-protective glass 
sphere that houses a 10-inch Hamamatsu photomultiplier tube together with 
electronic boards used for detection, digitization, and readout.  The strings 
are separated by an average distance of 125\,m, each one hosting 60 DOMs equally 
spaced over the kilometer of instrumented length.  The DOMs detect Cherenkov 
radiation produced by relativistic particles passing through the ice, including
muons and muon bundles produced by cosmic-ray air showers in the 
atmosphere above IceCube.  These atmospheric muons form a large background for 
neutrino analyses, but also provide us with an opportunity to use IceCube as a 
large cosmic-ray detector.  

In order to reject background signals produced by the 500\,Hz dark noise
from each DOM, a local coincidence in time with an interval of $\pm$ 1\,$\mu$s
is required between neighboring DOMs.  A trigger is then produced when eight 
or more DOMs detect photons in local coincidence within 5\,$\mu$s~\citep{Kelley:2014}.  
The trigger rate in IceCube, predominantly from atmospheric muons, ranges between 
2 and 2.4\,kHz.  This modulation is due to the large seasonal variation of the 
stratospheric temperature and consequently the density, which affects the decay 
rate of mesons into muons~\citep{Duperier:1949, Duperier:1951, Barrett:1952, 
Trefall:1955}.  The effect has also recently been studied with IceCube 
data~\citep{Tilav:2010jan, Desiati:2011aug}.  The detected muon events are 
generated by primary cosmic-ray particles with median energy of about 20\,TeV,
according to simulations.

The IceTop air-shower array~\citep{IceCube:2013feb} consists of 81 surface 
stations, with two light-tight tanks per station.  Each tank is 1.8\,m in 
diameter, 1.3\,m in height, and filled with transparent ice up to a height 
of 0.9\,m.  It hosts two DOMs, operating at different gains for an increased 
dynamic range.  The trigger in IceTop requires at least three stations 
to have recorded hits within a time window of 5\,$\mu$s~\citep{Kelley:2014}.
IceTop detects showers at a rate of approximately 30\,Hz with a minimum primary 
particle energy threshold of about 400\,TeV.  Its surface location near the
shower maximum makes it sensitive to the full electromagnetic component of the
shower, not just the muonic component.

Due to the limited satellite bandwidth for data transmission to the Northern
Hemisphere, IceCube data are analyzed online and reduced according to various
physics-motivated event selections.  All events that trigger IceCube are processed 
through two directional reconstruction procedures.  Their arrival direction is 
first estimated using a $\chi^2$ linear-track fit to the DOM hits.  Then, using 
this estimate as a seed, a more complex likelihood-based reconstruction is applied, 
accounting for aspects of light generation and propagation in the 
ice~\citep{IceCube:2014mar}.  The likelihood-based fit provides a median angular 
resolution of $3^{\circ}$ according to simulation~\citep{IceCube:2011oct}. 
This angular resolution, which lacks offline post-processing of event reconstructions 
and is therefore not typical of neutrino analyses in IceCube, worsens past zenith 
angles of approximately $70^{\circ}$.  The analysis is therefore limited to a 
declination range of $-90^{\circ} < \delta < -25^{\circ}$.  Note that at the South 
Pole, declination $\delta$ and zenith angle $\theta$ are directly related 
($\delta = \theta - 90^{\circ}$).  Simulation studies show that the direction of 
an air-shower muon is typically within 0.2$^{\circ}$ of the direction of the parent 
cosmic-ray particle~\citep{Abbasi:2013gh}, so the arrival direction distribution of
muons recorded in the detector is also a map of the primary cosmic-ray arrival
directions.
Due to the nature of cosmic-ray showers as a background for neutrino studies
and the limited data transfer rate available from the South Pole, all
cosmic-ray data are stored in a compact data storage and transfer (DST) format,
containing the results of the angular reconstructions described as well as
some limited information per event. 

Due to the limited transmission bandwidth, data collected by IceTop necessitate 
a prescale, which changed from year to year with growing detector configurations.
However, all showers that trigger eight or more stations are never prescaled and
thereby provide a consistent data set.  Only these events were used for this 
analysis, resulting in an event rate of about 1\,Hz and a high-energy data set 
with a median energy of 1.6\,PeV.  The angular resolution is a function of energy.  
The 68\% resolution is about $0.6^{\circ}$ at 1\,PeV and $0.3^{\circ}$ at 
10\,PeV~\citep{IceCube:2013feb}.

The experimental data used in this analysis were collected between May 2009 and
May 2015.  In the first two years, IceCube and IceTop operated in partial
detector configurations, with 59 active strings/stations (IC59/IT59) from May
2009 to May 2010, and 79/73 strings/stations (IC79/IT73) from May 2010 to May
2011.  The number of reconstructed events in IceCube and IceTop for each analysis
year are shown in Tab.\,\ref{tab:data} along with the corresponding detector
livetime in days and as a percentage that accounts for detector uptime and data
run selection.  It indicates the improved stability of the data sample over the
analysis period.  The table shows that in roughly 2062 days IceCube collected 
about 318 billion events and IceTop collected 170 million high-energy events.

The simulated data used in this paper were created using the standard air shower 
Monte Carlo program CORSIKA~\citep{Heck:1998vt}, the SIBYLL hadronic interaction 
model (Version 2.1)~\citep{Ahn:2009wx}, and a full simulation of the IceCube
and IceTop detectors.  For the primary cosmic-ray composition and energy spectrum, 
we assume a mixed model based on~\citet{Hoerandel:2002yg}.

\section{Analysis}
\label{sec:analysis}

\subsection{Method}
\label{subsec:method}

The analysis methods for this work have been published previously
in~\citet{IceCube:2011oct}; what follows is a brief overview. 
All sky maps shown were made using HEALPix~\citep{Gorski:2005apr}, 
a mapping program that pixelizes the sky into bins of equal solid-angle. 
For this work, a pixel size of approximately $(0.84^{\circ})^2$ 
($N_\mathrm{side} = 64$) is used.

In order to study the anisotropy, we need to compare the actual sky map of 
cosmic-ray arrival directions (``data map'') to a sky map which represents
the response of the detector to an isotropic cosmic-ray flux (``reference 
map'').  Due to detector effects, for example nonuniform exposure to different 
parts of the sky and gaps in the detector uptime, the reference map is not
itself isotropic.  The reference map can be determined by integrating the 
time dependent exposure of the detector over the livetime.  We determine the
exposure from the data themselves using the time-scrambling method described 
in~\citet{Alexandreas:1993may}, a standard method in the search for gamma-ray, 
cosmic-ray, and neutrino sources for large-field-of-view detectors.  In brief, 
for each detected event stored in the data map, 20 “fake” events are generated 
by keeping the local zenith and azimuth angles fixed and calculating new values 
for right ascension using times randomly selected from within a pre-defined 
time window $\Delta t$ bracketing the time of the event being considered.  
These fake events are stored in the reference map with a weight of 1/20.  
The creation of several ``fake'' events per real event and subsequent weighting 
serves to reduce statistical fluctuations.  The size of the time window 
$\Delta t$ determines the sensitivity of the search to features of 
various angular sizes: a time window of four hours would make a search 
sensitive to structures of 4\,hr / 24\,hr $\times$ 360$^{\circ}$ = 60$^{\circ}$ 
or smaller in size.  In this work, a scrambling period of 24 hours is used to 
make the search sensitive to structures on all angular scales.  The choice of 
24 hours scrambling time is possible because the local arrival direction 
distribution, i.e., the distribution of zenith and azimuth angles in local 
detector coordinates, is stable within such a time interval (see 
Section~\ref{sec:systematics}).
\begin{figure}[t]
  \centering
    \includegraphics[width=\columnwidth]{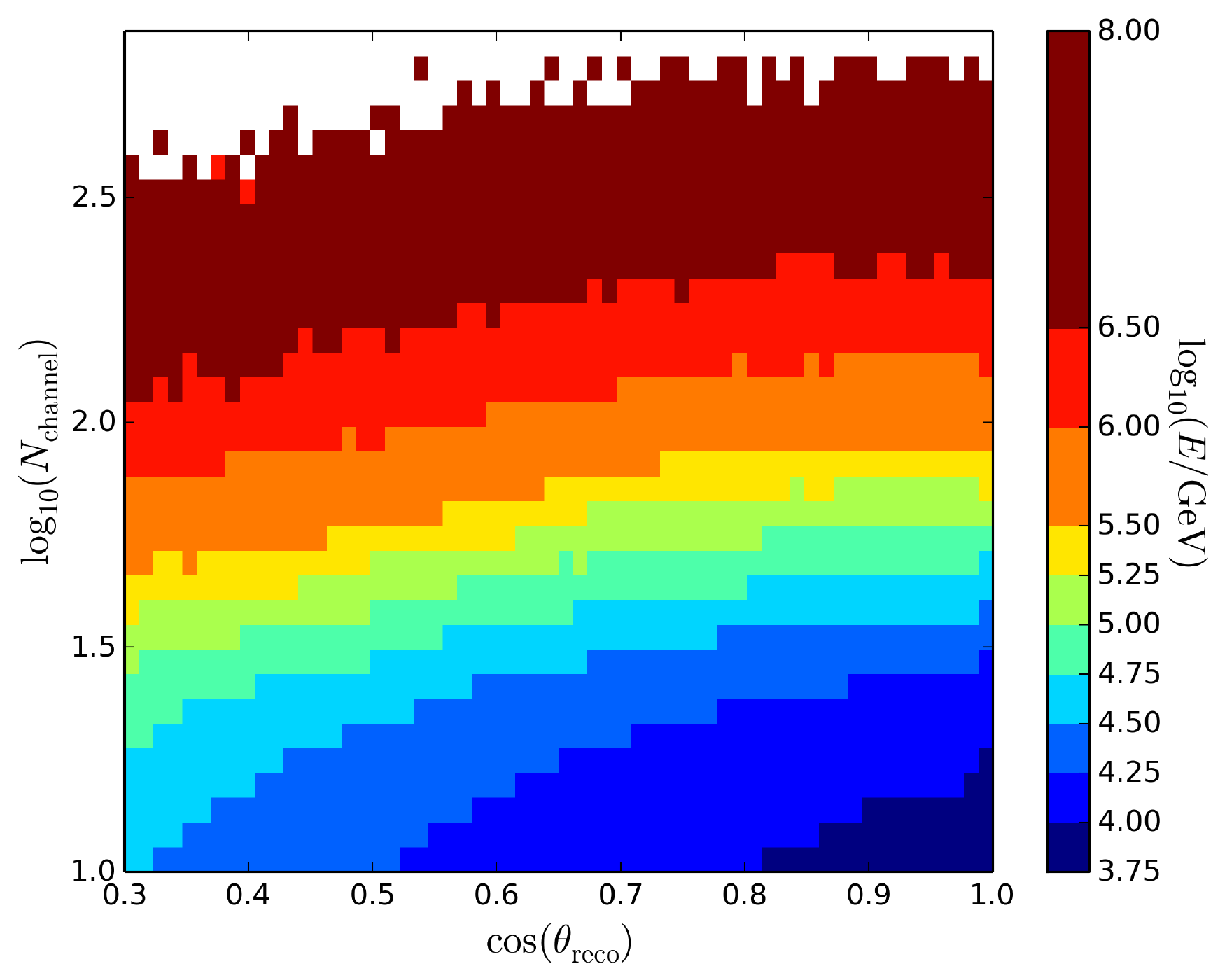}
    \caption{Median true energy as a function of the cosine of the reconstructed
zenith angle $\theta_\mathrm{reco}$ and the number of DOMs hit in the event,
$N_\mathrm{channel}$, from simulation.}
    \label{fig:med_energy}
\end{figure}

It is important to emphasize that scrambling the time of events with a given 
zenith angle in local detector coordinates is equivalent to randomly modifying 
the right ascension of the event within the same declination band, with the 
width of the band determined by the pixelization used.  As a result, the 
residual between the actual arrival direction distribution and the reference 
maps, determined by independently normalizing each declination band, is sensitive 
primarily to anisotropy modulations in right ascension~\citep{Ahlers:2016njl}.
Simulation studies~\citep{Santander:2013} indicate that any structure is 
effectively reduced to its projection onto right ascension, limiting the
sensitivity to the determination of the true anisotropy.  The relative
intensity of any small-scale features is also underestimated because of the
overestimation of the isotropic ``floor'' for the entire declination band.
This distortion is unavoidable, but for the small level of anisotropy and the
choice of 24 hours scrambling time in this analysis, the effect is not 
significant.
\begin{figure*}[ht]
  \centering  
    \includegraphics[width=0.49\textwidth]{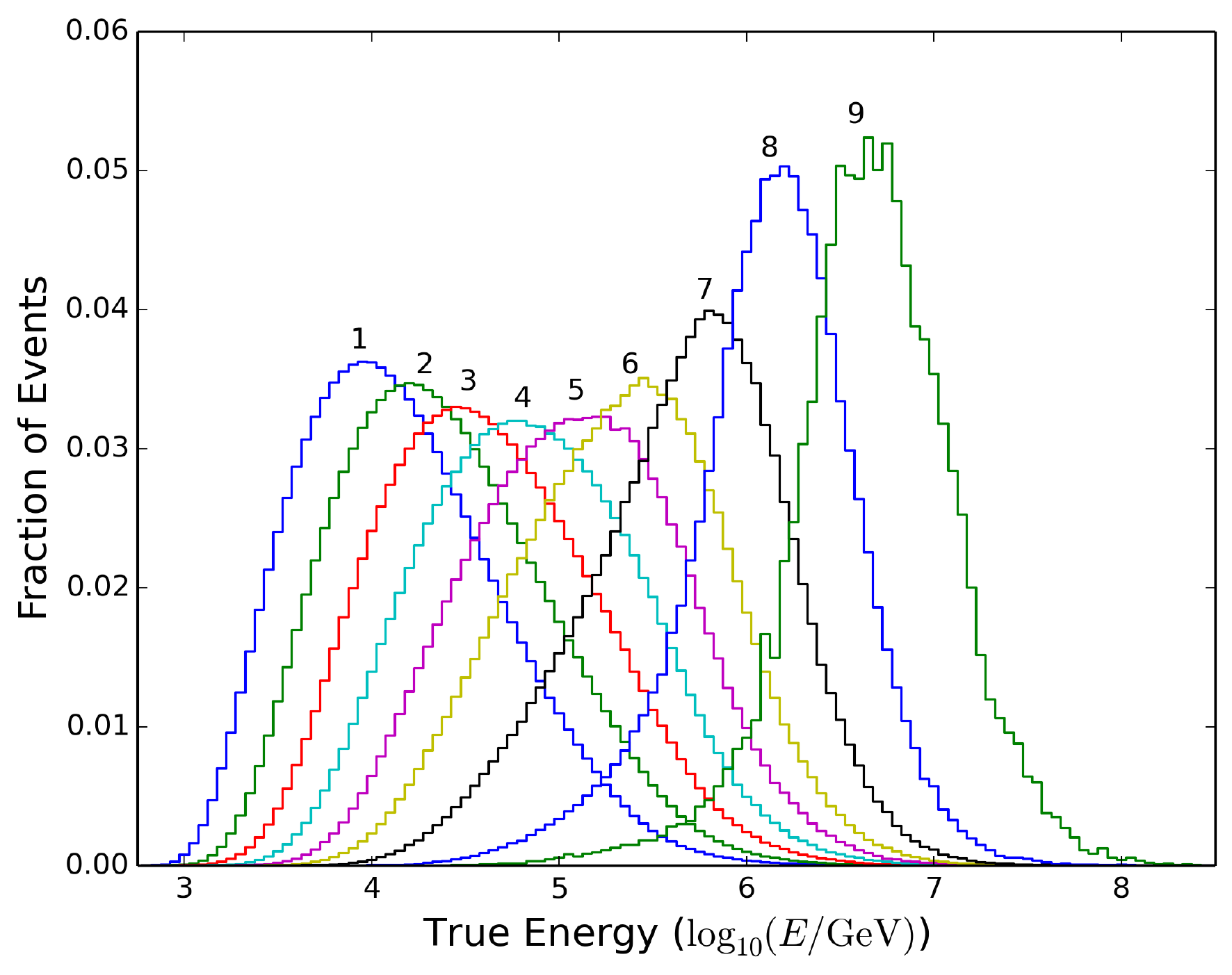}
    \includegraphics[width=0.49\textwidth]{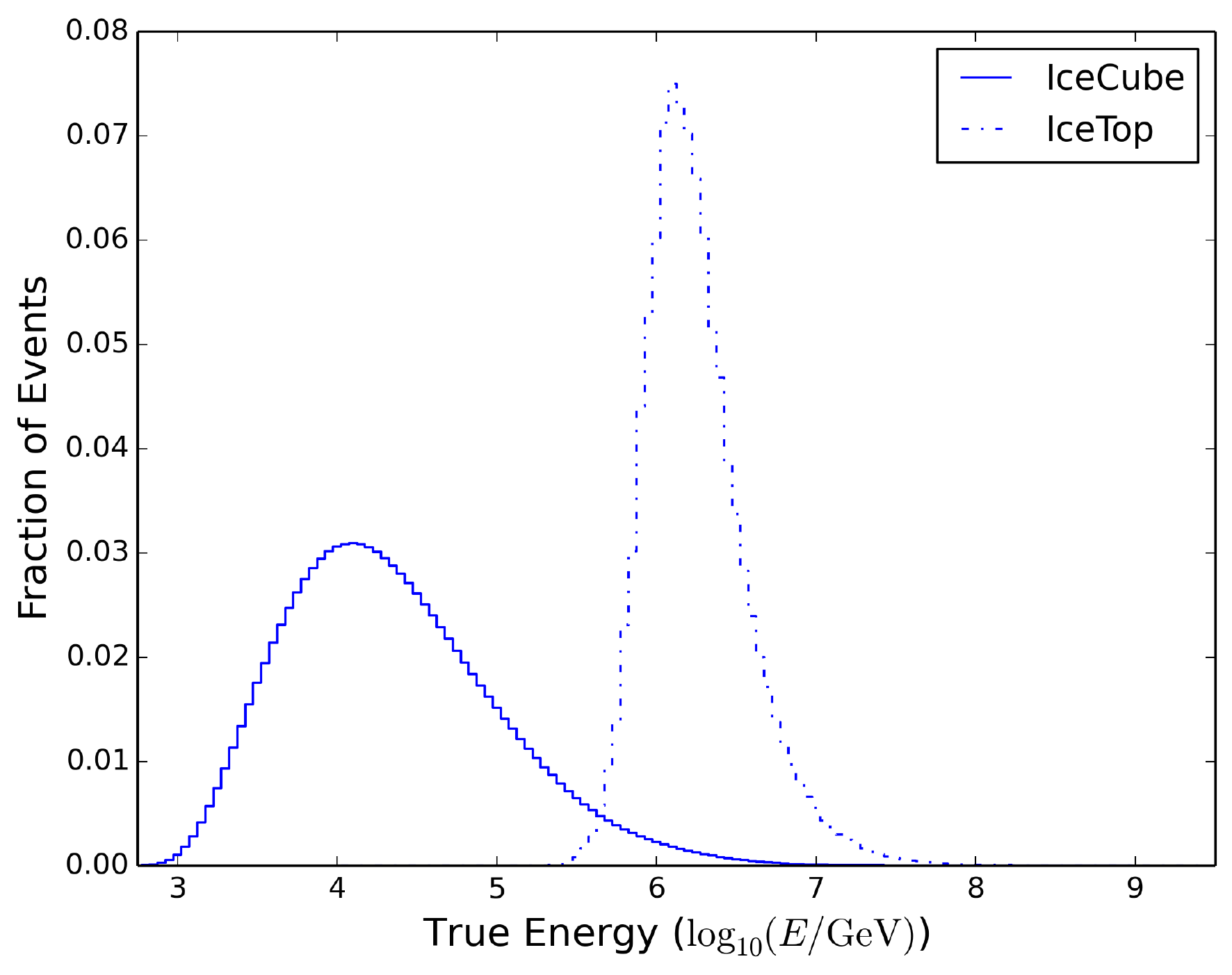}
    \caption{\textit{Left:} Fraction of events as a function of the true energy
for the nine energy bins used in this analysis, from simulation.  The median 
energy and the 68\% central interval for the nine bins are listed in the first
row of Tab.\,\ref{tab:proj}.  The primary cosmic-ray composition and energy 
spectrum are based on~\citet{Hoerandel:2002yg}. \textit{Right:} Fraction of events
as a function of true energy for the full IceCube (solid line) and IceTop (dashed
line) data sets used in this analysis.  The IceCube data set includes the data from
all nine energy bins plus low energy events not included in the lowest energy
bin (see text).}
    \label{fig:energy_dist}
\end{figure*}

The map of cosmic-ray anisotropy is obtained by calculating the residual
between the data map and the reference map.  The relative intensity is defined 
as $\delta I_i = (N_i - \langle N \rangle_i)/ \langle N \rangle_i$, where 
$N_i$ and $\langle N \rangle_i$ are the number of observed events and the 
number of reference events in the $i^{th}$ pixel, respectively.  Maps showing 
the statistical significance of deviations are calculated according 
to~\citet{LiMa:1983sep}.  To study the small-scale anisotropy, the dipole 
and quadrupole terms of the spherical harmonic functions were fit to the 
data and then subtracted.  All maps undergo a top-hat smoothing procedure 
in which a single pixel's value is the sum of all pixels within a given 
angular distance, or smoothing radius.  In the case of this data set, the 
median angular resolution, as found from simulation, is 3$^{\circ}$. 
Therefore, a smoothing of $5^{\circ}$, roughly equivalent to the optimal 
bin size for point-source searches~\citep{Alexandreas:1993may}, is applied 
to the maps.

\subsection{Separation into Energy Bins}
\label{subsec:energy}

In order to study the anisotropy in cosmic-ray arrival direction as a function
of primary energy in IceCube, a similar energy estimation procedure to that
of~\citet{IceCube:2012feb} is used.  Events are classified using the number of 
DOMs that detected Cherenkov light, $N_\mathrm{channel}$, and the reconstructed 
zenith angle, $\theta_\mathrm{reco}$.  $N_\mathrm{channel}$ is used as an energy 
estimator of the muons detected by IceCube.  The reconstructed angle 
$\theta_\mathrm{reco}$ is considered because at larger zenith angles muons, and 
therefore the primary cosmic-ray particles, must have higher energy in order to 
reach and trigger the buried IceCube experiment.  Simulation data are used to 
determine bands in primary particle energy as a function of $N_\mathrm{channel}$ 
and the cosine of $\theta_\mathrm{reco}$, as shown in Fig.\,\ref{fig:med_energy}.
The figure shows that for a given $N_{\mathrm{channel}}$, events at larger zenith 
angles are produced by cosmic-ray particles with higher energy.
\begin{table*}[!ht]
  \centering
  \begin{tabular}{cccc}
    \hline
    Region  &  Right Ascension (deg) & Declination (deg) & Peak Significance \\
    \hline
    1a & $142.5_{-2.4}^{+4.9}$ & $-49.7_{-3.9}^{+2.3}$ & $11.0\sigma$ \\
    1b & $110.5_{-3.5}^{+5.3}$ & $-55.9_{-2.3}^{+5.4}$ & $6.9\sigma$ \\
    2  & $261.0_{-8.5}^{+3.4}$ & $-48.9_{-2.3}^{+4.7}$ & $11.4\sigma$ \\
    3  & $200.4_{-1.4}^{+2.8}$ & $-38.7_{-2.3}^{+2.3}$ & $10.8\sigma$ \\
    4  & $327.9_{-16.8}^{+11.9}$ & $-74.6_{-4.4}^{+4.4}$ & $11.0\sigma$ \\
    5  & $215.6_{-8.6}^{+18.7}$ & $-72.4_{-2.2}^{+5.2}$ & $-9.3\sigma$ \\
    6  & $74.5_{-4.2}^{+4.2}$ & $-36.4_{-3.8}^{+5.0}$ & $-10.3\sigma$ \\
    7a & $317.1_{-2.1}^{+4.2}$ & $-38.7_{-1.5}^{+5.2}$ & $-7.2\sigma$ \\
    7b & $292.5_{-1.4}^{+1.4}$ & $-41.0_{-1.6}^{+1.6}$ & $-9.6\sigma$ \\
    8  & $164.7_{-2.9}^{+3.2}$ & $-48.1_{-3.9}^{+4.7}$ & $-11.9\sigma$ \\
    $9^{*}$  & $94.1_{-40.9}^{+9.4}$ & $-82.0_{-2.2}^{+5.1}$ & $-7.9\sigma$ \\
    $10^{*}$ & $27.4_{-1.4}^{+4.9}$ & $-27.3_{-2.0}^{+3.3}$ & $10.6\sigma$ \\
    \hline
  \end{tabular}
  \caption{Locations and pre-trial peak significance values for the small-scale
structures visible after subtracting the best-fit dipole and quadrupole
functions.  Errors indicate the positions of the farthest pixels within $1\sigma$
of the peak significance.  Regions marked {\it a} and {\it b} were previously reported 
as one region.  Regions with an asterisk are new to this analysis.}
  \label{tab:minmax}
\end{table*}
\begin{figure*}[ht]
  \centering
  \includegraphics[width=0.49\textwidth]{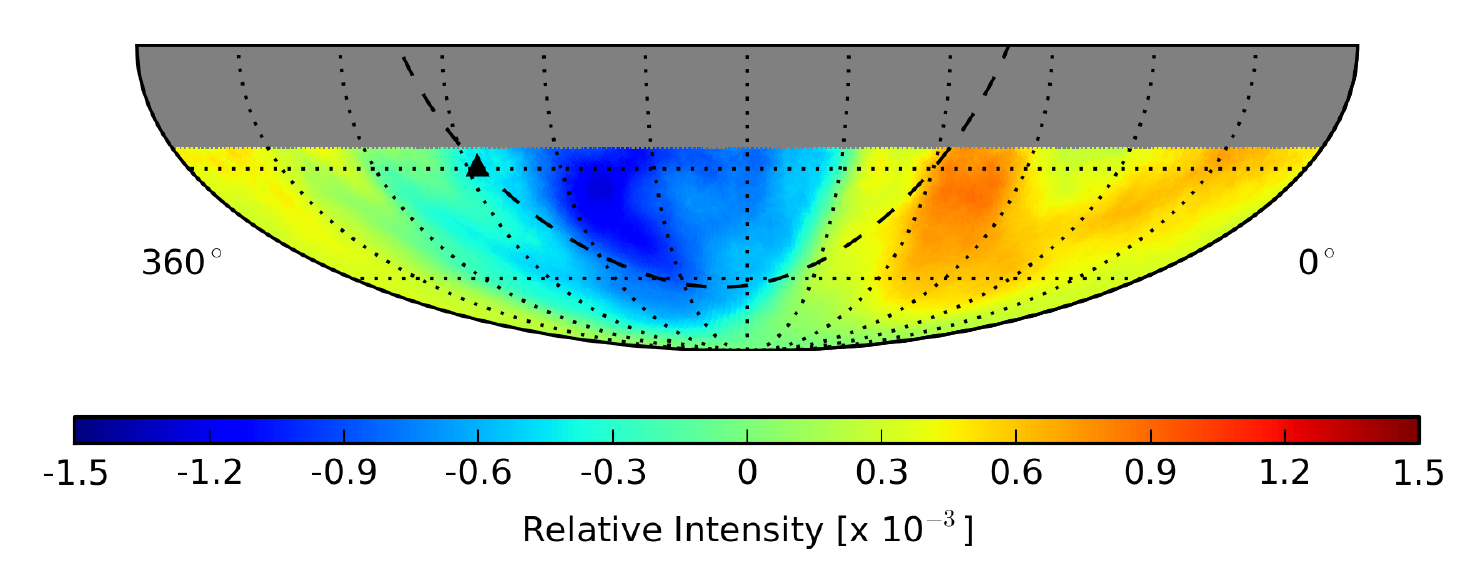}
  \includegraphics[width=0.49\textwidth]{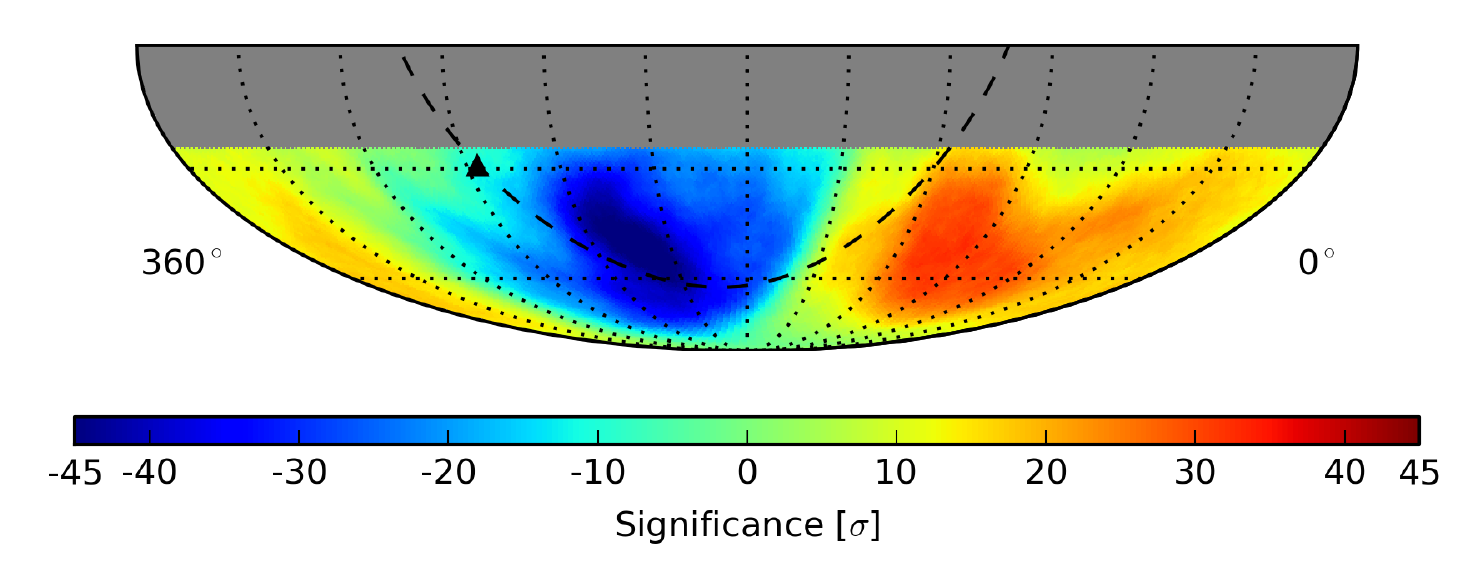}
  \includegraphics[width=0.49\textwidth]{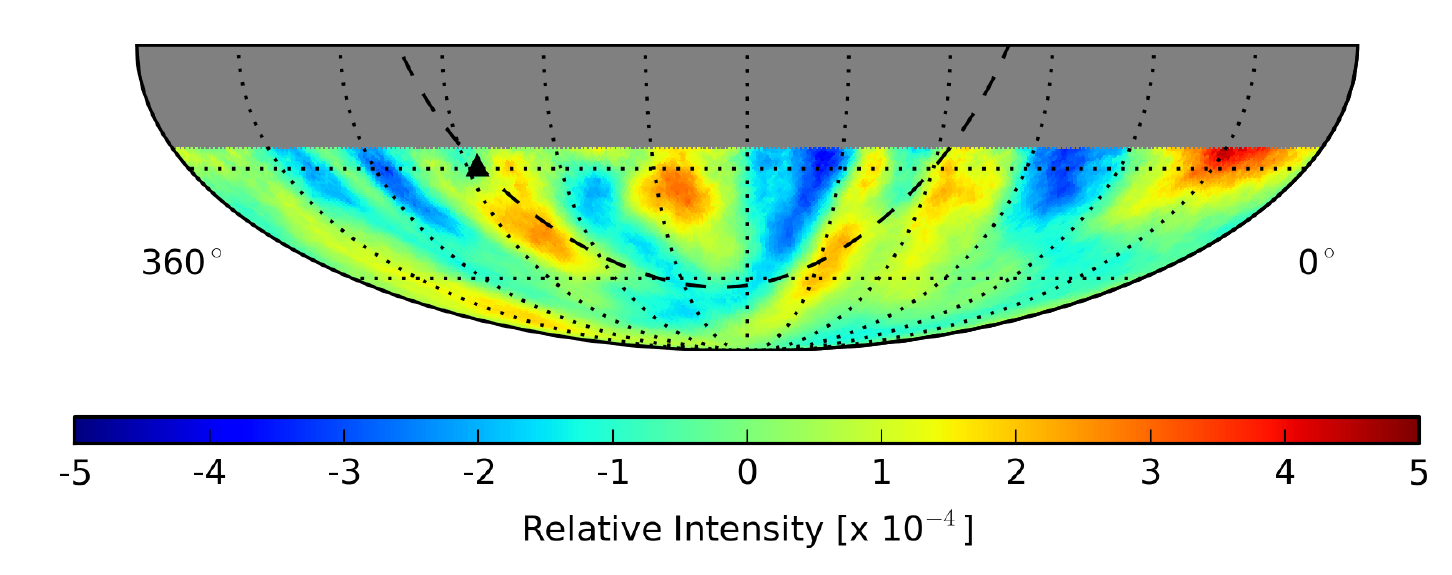}
  \includegraphics[width=0.49\textwidth]{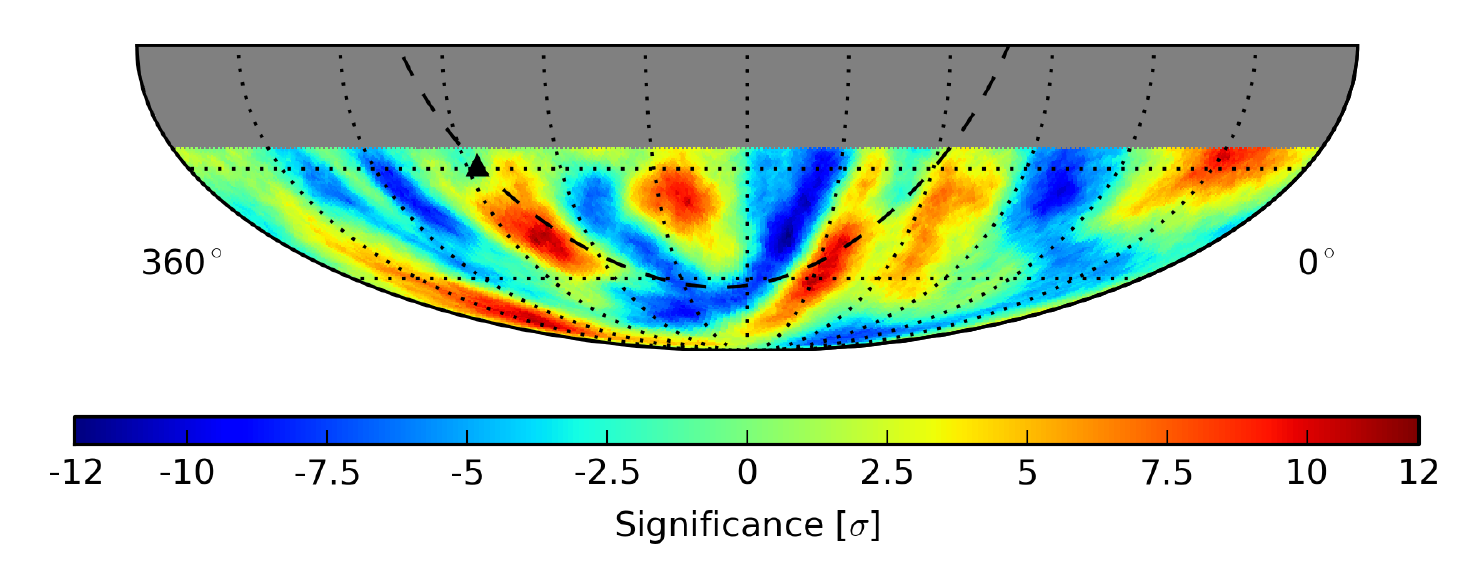}
  \caption{Relative intensity (\textit{left}) and pre-trial statistical
significance (\textit{right}) maps shown before (\textit{top}) and after
(\textit{bottom}) dipole- and quadrupole-subtraction. The maps are in
equatorial coordinates and use an angular smoothing radius of $5^{\circ}$.
The dashed line indicates the Galactic plane and the triangle indicates the
Galactic center.}
  \label{fig:largesmall}
\end{figure*}
A B-spline function (see \citet{Whitehorn:2013nh} for a description of the method 
applied here) in $N_\mathrm{channel}$ and $\cos$($\theta_\mathrm{reco}$) is used 
to fit the data in Fig.\,\ref{fig:med_energy} in order to reduce the errors due to 
limited simulation data statistics.  Events are then separated into nine energy bins 
with increasing mean primary particle energies ranging from 13\,TeV (bin 1) to 
5.4\,PeV (bin 9).  The resolution of this energy assignment depends on the detector 
configuration and energy band but is on the order of 0.5 in $\log_{10}(E/\mathrm{GeV})$.
It is primarily limited by the relatively large fluctuations in the fraction 
of the total shower energy that is transferred to the muon bundle.  
The individual colors in Fig.\,\ref{fig:med_energy} correspond to the different 
energy bins.  The dark blue bin corresponds to events with
$\log_{10}(E/\mathrm{GeV})<4.0$.  This bin is not used because of
the limited zenith range of the events.

The nine data samples created by this method are statistically independent, but
as a result of the limited energy resolution, the energy distributions of the
events in the bins overlap considerably.  To illustrate this, the left panel of 
Fig.\,\ref{fig:energy_dist} shows the fraction of events as a function of the true
primary energy for the nine energy bins, from simulation.  The median energy and 
the 68\% central interval for the nine bins are also listed in the first row of
Tab.\,\ref{tab:proj}.  The right panel of Fig.\,\ref{fig:energy_dist} shows the 
fraction of events as a function of the true primary energy for the entire data 
set, without energy cuts.  Note that this includes all events from the nine energy
bins, but also the events in the dark blue bin ($\log_{10}(E/\mathrm{GeV})<4.0$) of
Fig.\,\ref{fig:med_energy} which are not part of the lowest energy bin.  

In contrast to IceCube, the IceTop data set at present cannot be split into 
several energy bins because the total number of events is still too small.
For this work, a single high-energy IceTop sample was selected by using all
events with $N_\mathrm{station}\geq 8$, resulting in a median energy of 
1.6\,PeV.  Fig.\,\ref{fig:energy_dist} (\textit{right}) shows the fraction of 
events as a function of the true primary energy for the IceTop data set used in 
this analysis.  

\section{Results}
\label{sec:results}

\subsection{Large- and Small-Scale Structure}
\label{subsec:struct}

Figure\,\ref{fig:largesmall} shows relative intensity and pre-trial significance
sky maps for large- and small-scale structures.  All of the maps contain six years
of IceCube data at all energies.  The energy distribution of these events, i.e., the
fraction of events as a function of the true energy, is shown in the right panel of
Fig.\,\ref{fig:energy_dist}.  The maps are top-hat smoothed with a $5^{\circ}$ angular 
radius.  The relative intensity map, shown in the top left plot of Fig.\,\ref{fig:largesmall}, 
is similar to previously published work based on IC59 data~\citep{IceCube:2011oct} and shows 
anisotropy at the $10^{-3}$ level, characterized by a large excess from $30^{\circ}$ 
to $120^{\circ}$ and a deficit from $150^{\circ}$ to $250^{\circ}$.  The corresponding 
significance of the large-scale structure, shown in the top right plot of
Fig.\,\ref{fig:largesmall}, shows the increasingly high statistical significance
of the observation.  While this large-scale structure dominates the anisotropy,
there is also anisotropy on smaller scales.  This structure, with a relative
intensity on the order of $10^{-4}$, becomes visible after the best-fit dipole
and quadrupole are subtracted from the sky map.  The lower left panel of 
Fig.\,\ref{fig:largesmall} shows the relative intensity of the residual map,
and the lower right panel the corresponding significance map.  The maps show the
presence of cosmic-ray anisotropy at angular scales approaching the angular 
resolution of IceCube for cosmic-ray primaries.

\begin{figure*}[ht]
  \centering
  \includegraphics[width=0.49\textwidth]{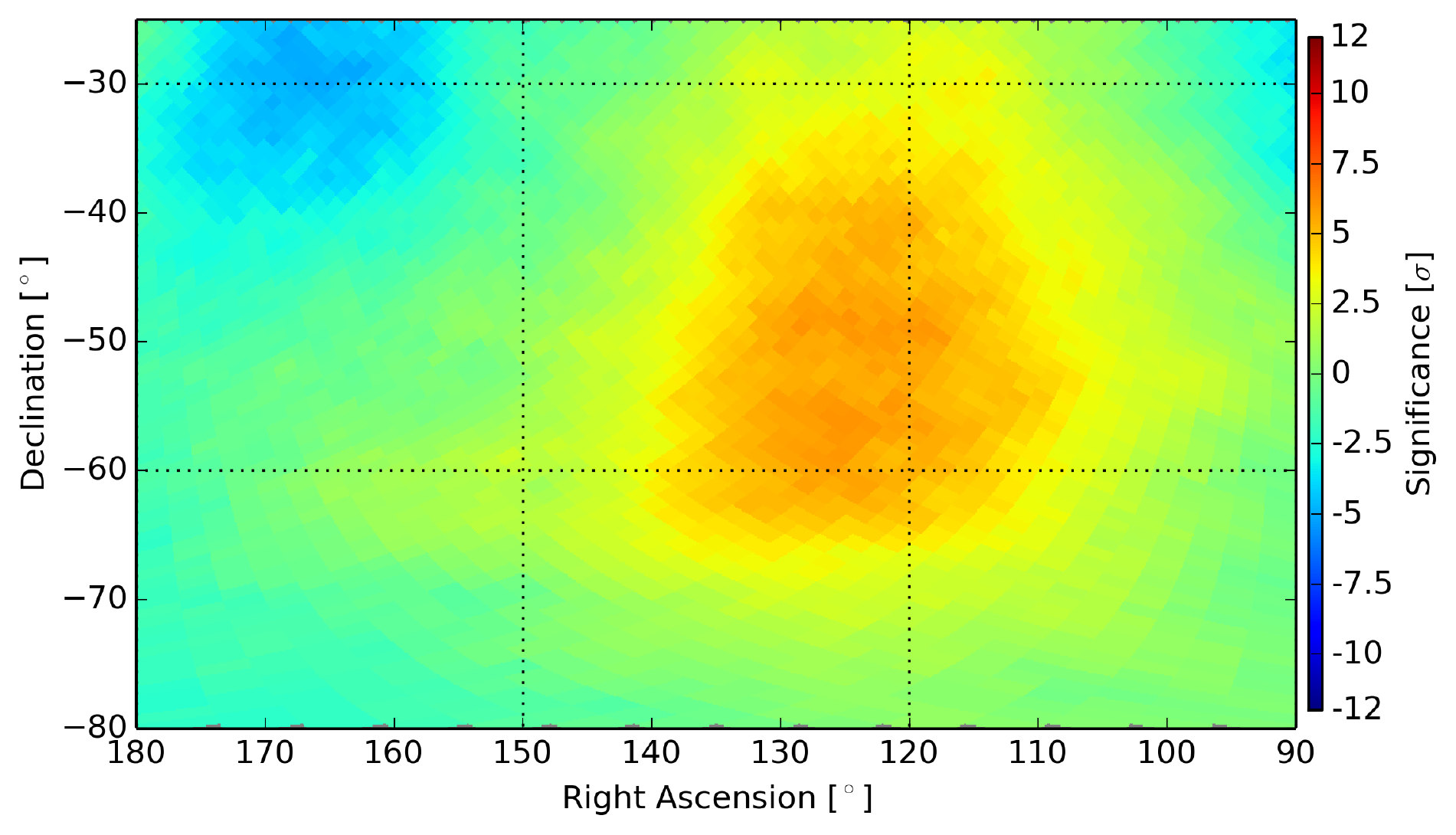}
  \includegraphics[width=0.49\textwidth]{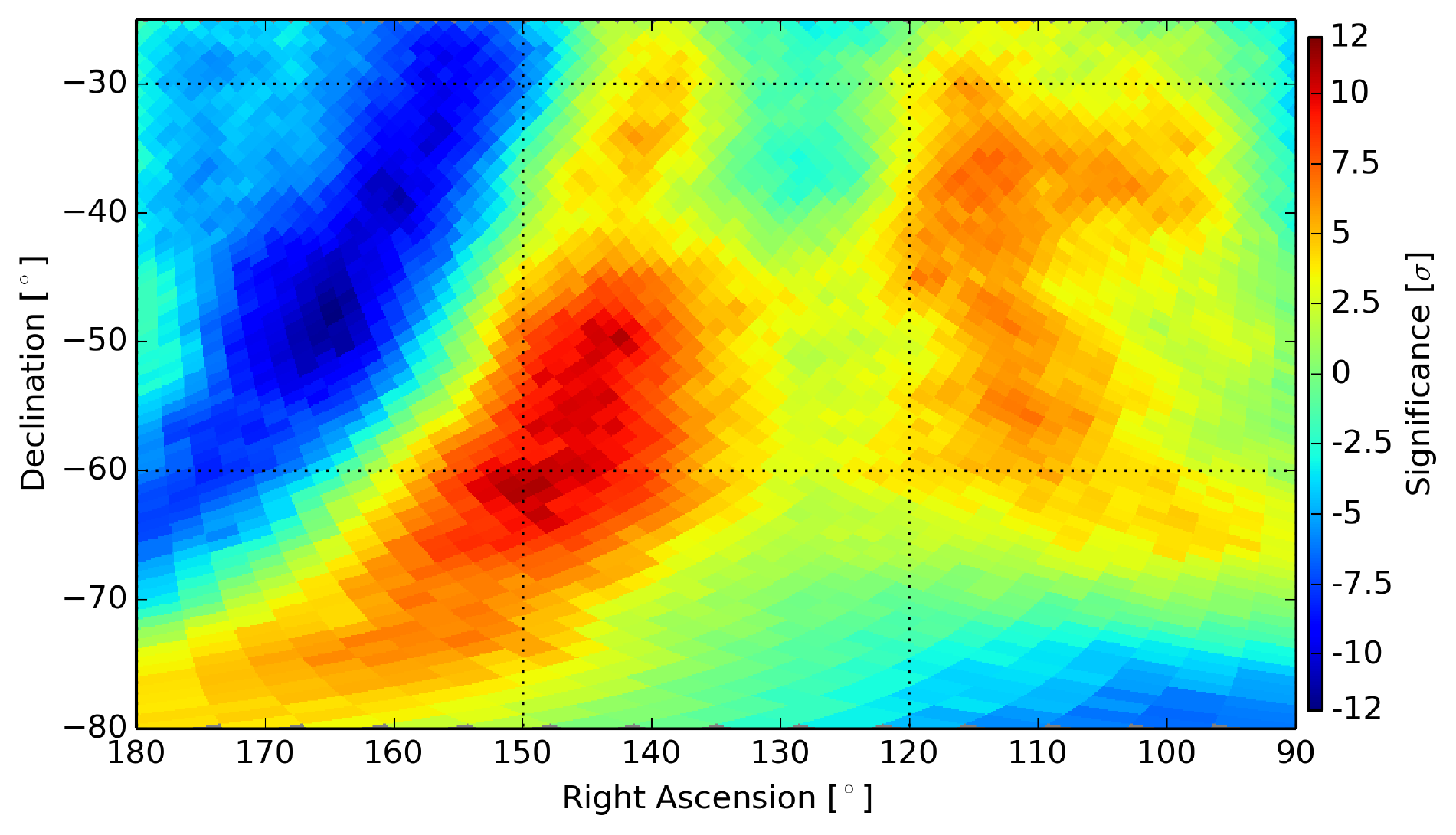}
  \caption{Significance map in the vicinity of Region 1 (see 
Tab.\,\ref{tab:minmax}) as previously published using only data taken with the IC59 
configuration~\citep{IceCube:2011oct} with $20^{\circ}$ smoothing (\textit{left}) 
and for the full data set used in this analysis with $5^{\circ}$ smoothing
(\textit{right}). Maps are shown in equatorial coordinates.}
  \label{fig:icsquare}
\end{figure*}
\begin{figure*}[ht]
  \centering
    \includegraphics[width=0.8\textwidth]{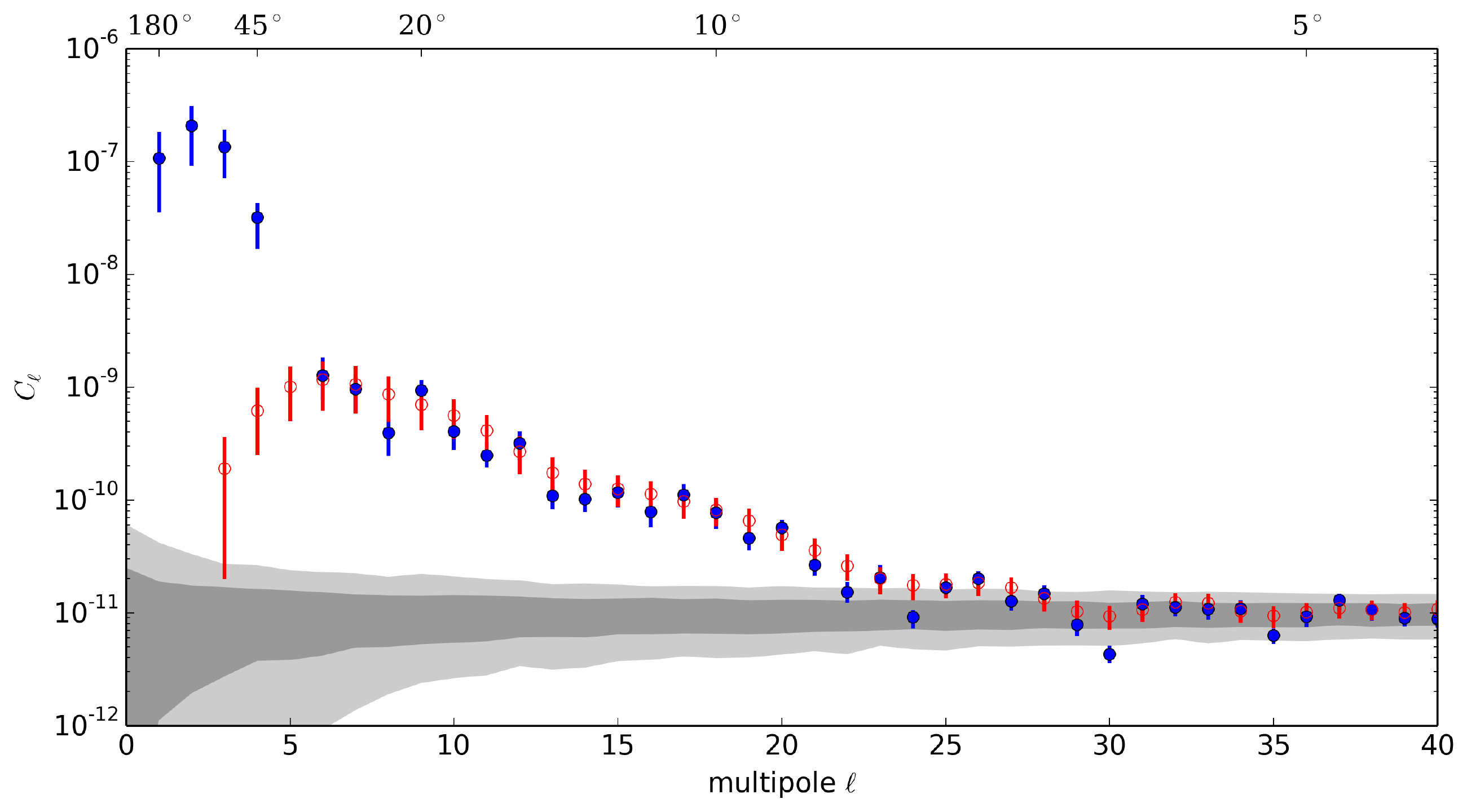}
    \caption{Angular power spectra for the relative intensity map for 
six years of IceCube data.  Blue and red points show the power spectrum before and 
after the subtraction of the best-fit dipole and quadrupole terms from the relative 
intensity map.  Error bars are statistical (see the text for a discussion of 
systematic errors).  The gray bands indicate the 68\% (dark) and 95\% (light) 
spread in the $C_{\ell}$ for a large sample of isotropic data sets.  The power
spectrum is calculated using the unsmoothed map.}
    \label{fig:powspec}
\end{figure*}

Table\,\ref{tab:minmax} shows the positions and peak significances of excess and 
deficit regions with a pre-trial significance exceeding $5\sigma$.  The regions are 
numbered to maintain consistency with~\citet{IceCube:2011oct} whenever possible.
The significances quoted are pre-trial, and any blind search would have to account
for the fact that we search for significant excess or deficit regions anywhere
in the roughly $10^4$ independent bins of the map.  However, all but two of the regions
listed in Tab.\,\ref{tab:minmax} have been previously reported in the analysis of 
the IC59 data set.  In the new data set, which includes the IC59 data set but is a 
factor of nine larger, all regions appear with greatly increased significance.

Figure\,\ref{fig:icsquare} provides an example of how the high-statistics data set 
reveals more details of the anisotropy.  The left plot shows a region of excess using 
only the IC59 data set with $20^{\circ}$ smoothing applied, as 
in~\citet{IceCube:2011oct}.  The same region is shown in the right plot, using
the updated data set and $5^{\circ}$ angular smoothing.  What previously appeared
as one region is now observed as two distinct regions, each at high significance.  
This difference does not appear to be the result of a time dependence of the 
small-scale structure, as the same split is visible in the IC59 map with $5^{\circ}$ 
smoothing but not at high enough significance to be previously reported. 

\begin{figure*}[ht]
  \centering
  \includegraphics[width=0.49\textwidth]{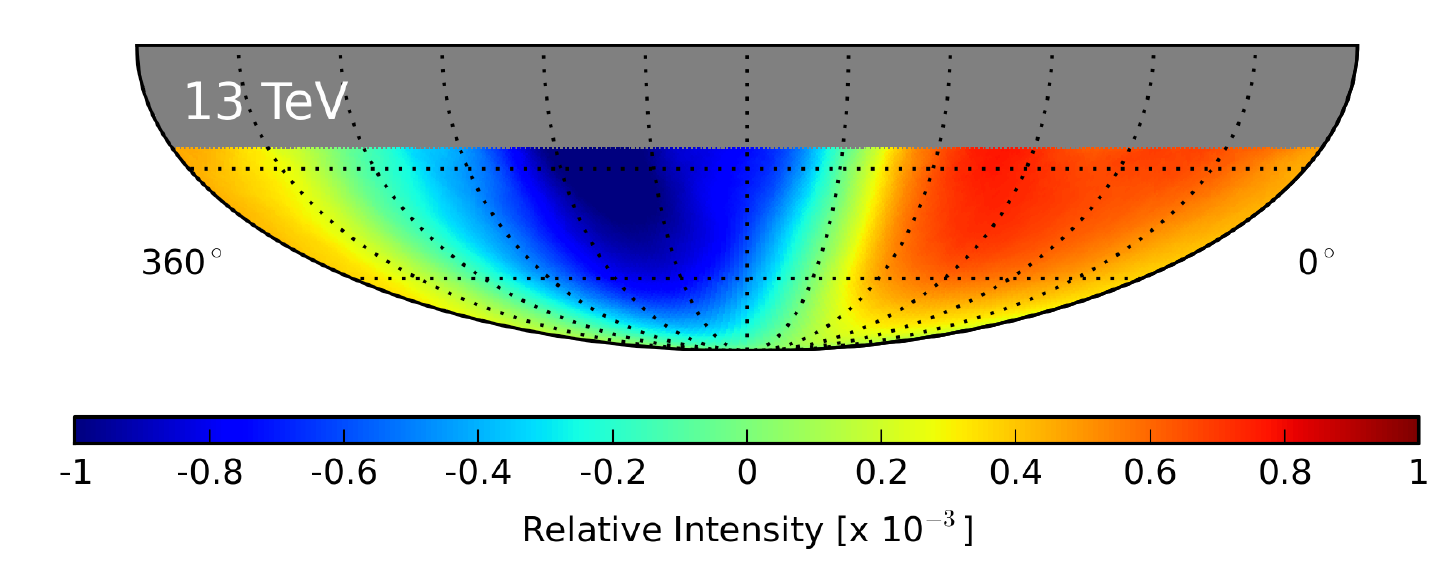}
  \includegraphics[width=0.49\textwidth]{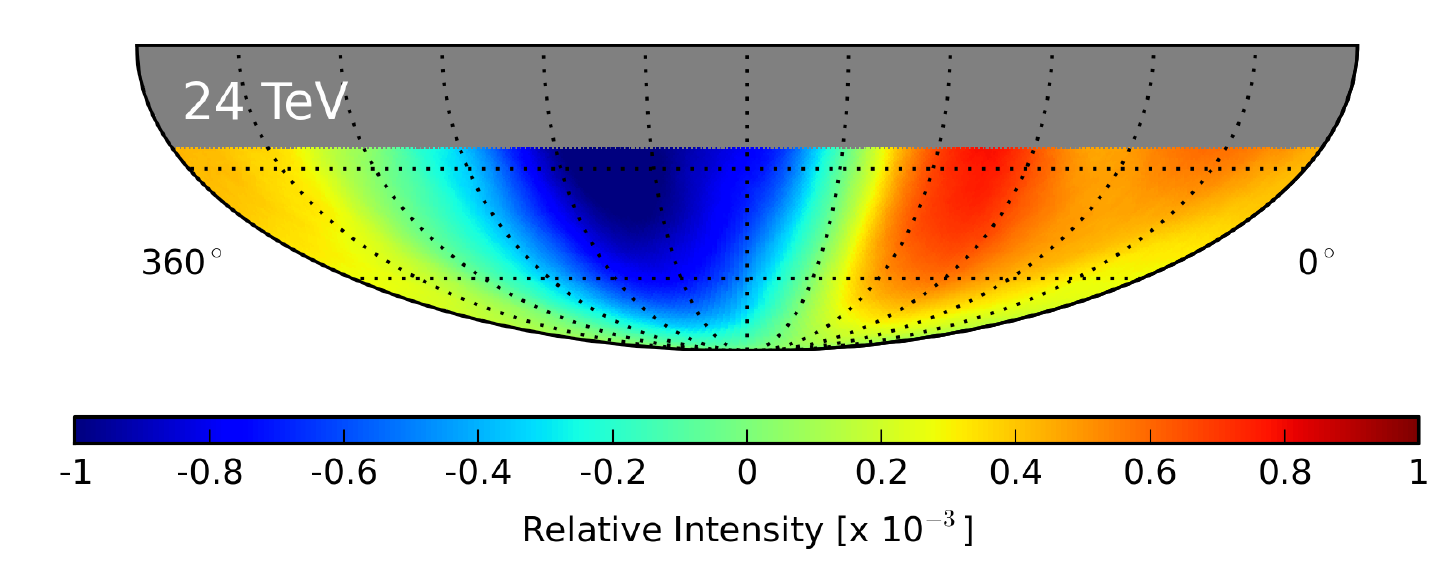}
  \includegraphics[width=0.49\textwidth]{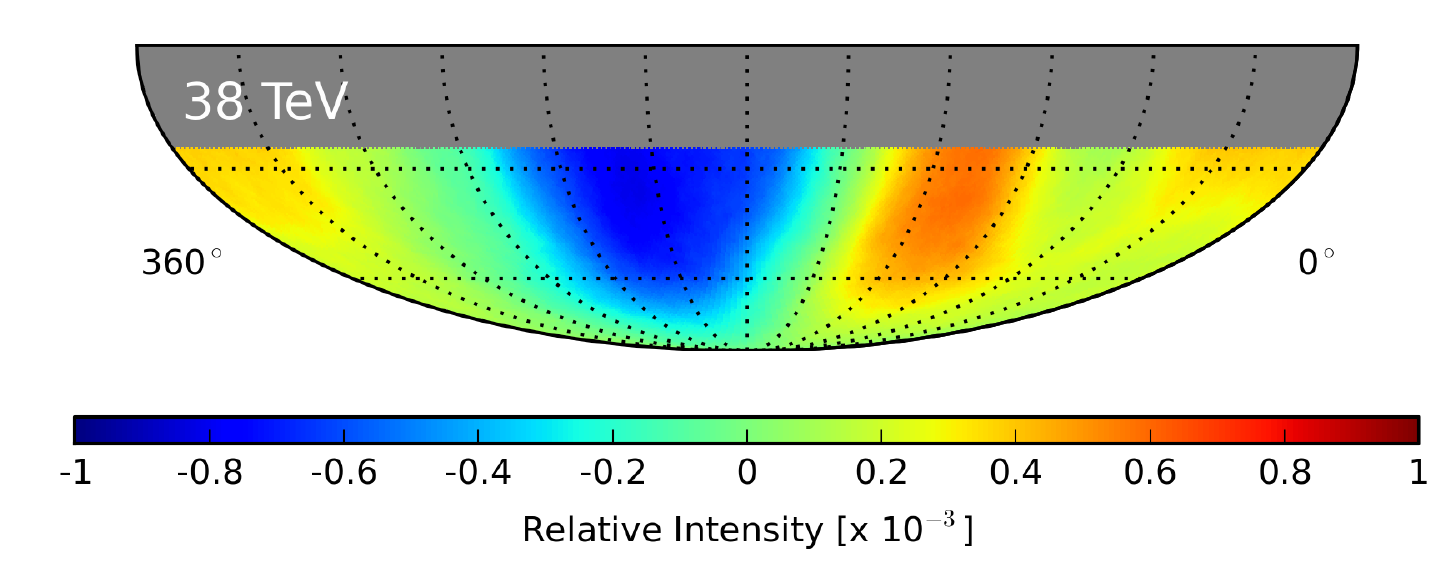}
  \includegraphics[width=0.49\textwidth]{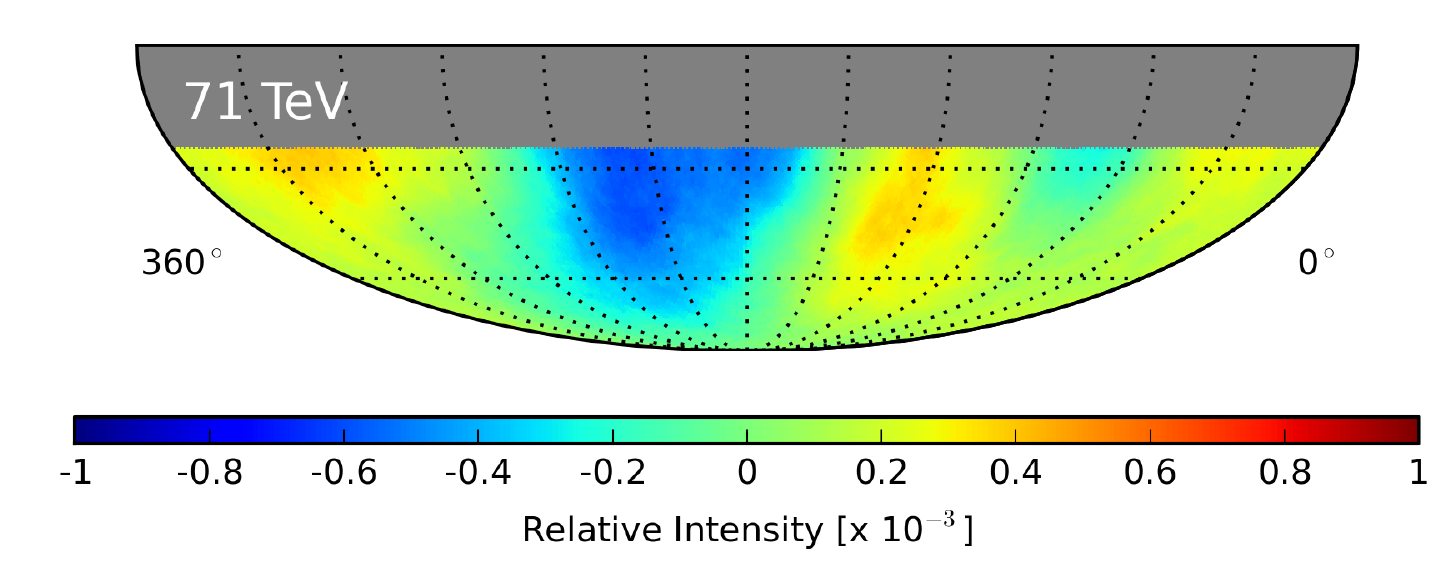}
  \includegraphics[width=0.49\textwidth]{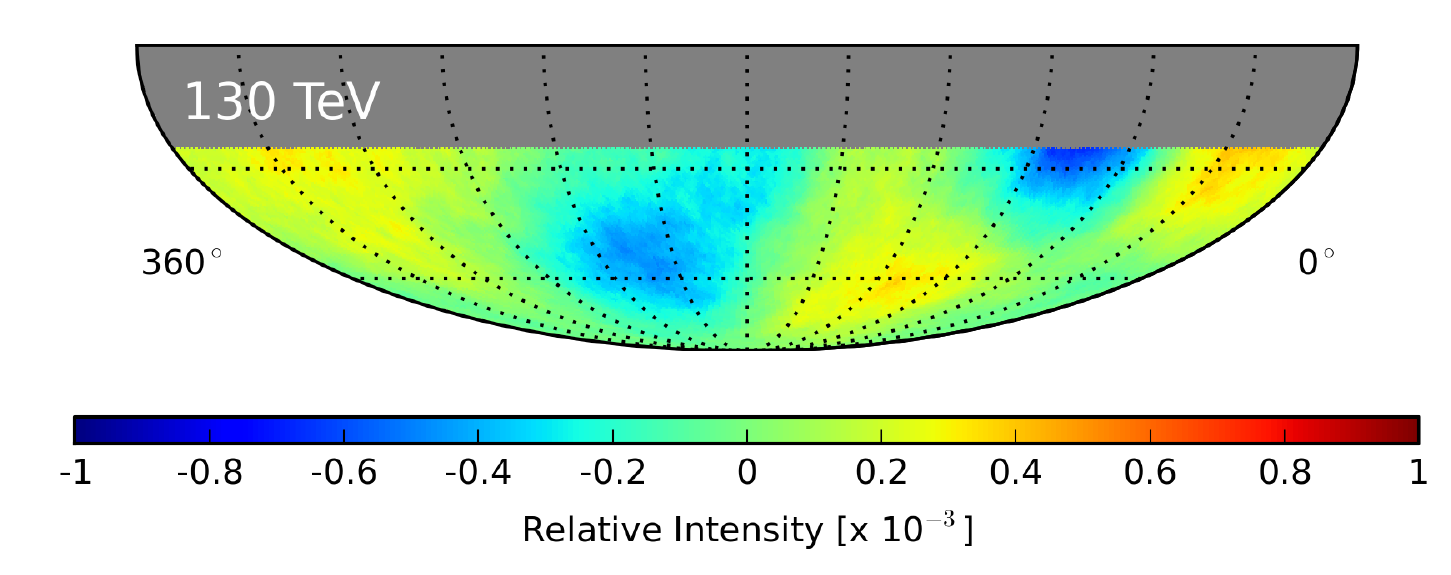}
  \includegraphics[width=0.49\textwidth]{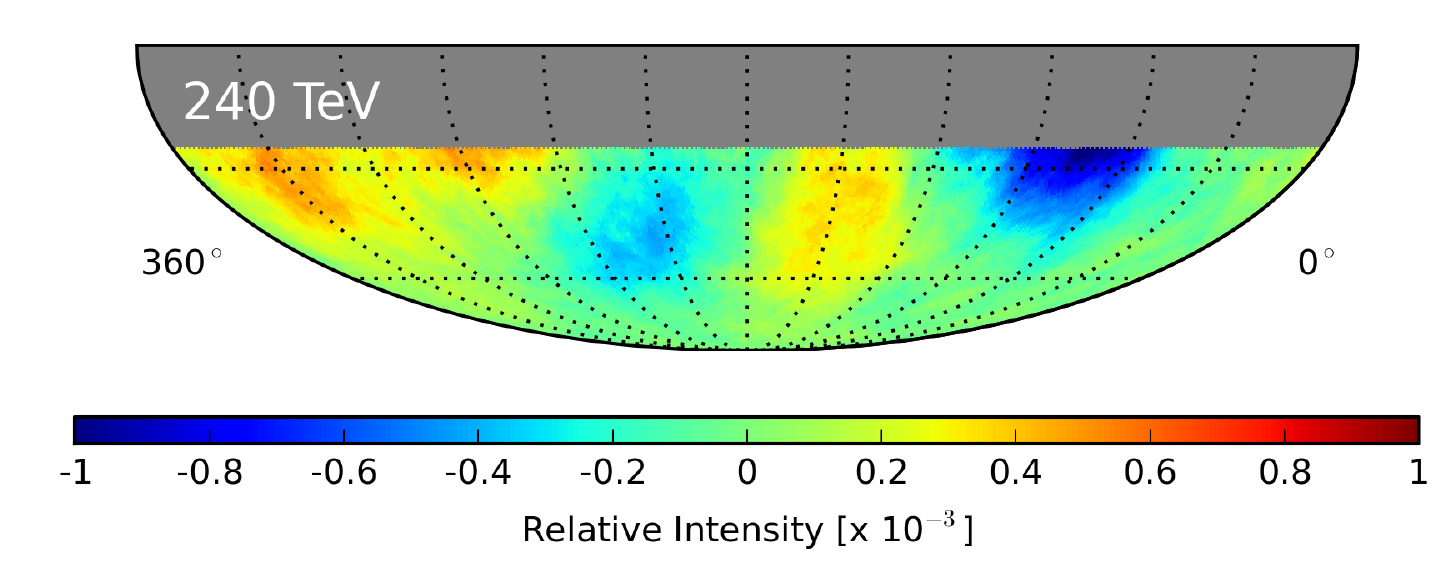}
  \includegraphics[width=0.49\textwidth]{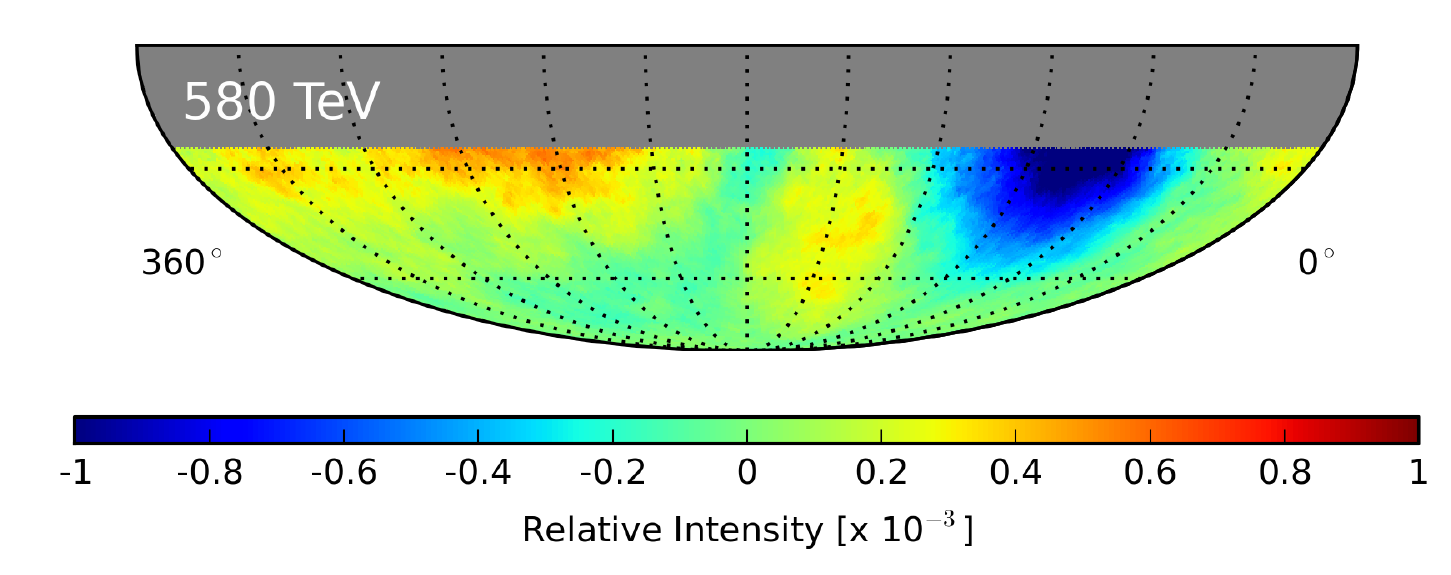}
  \includegraphics[width=0.49\textwidth]{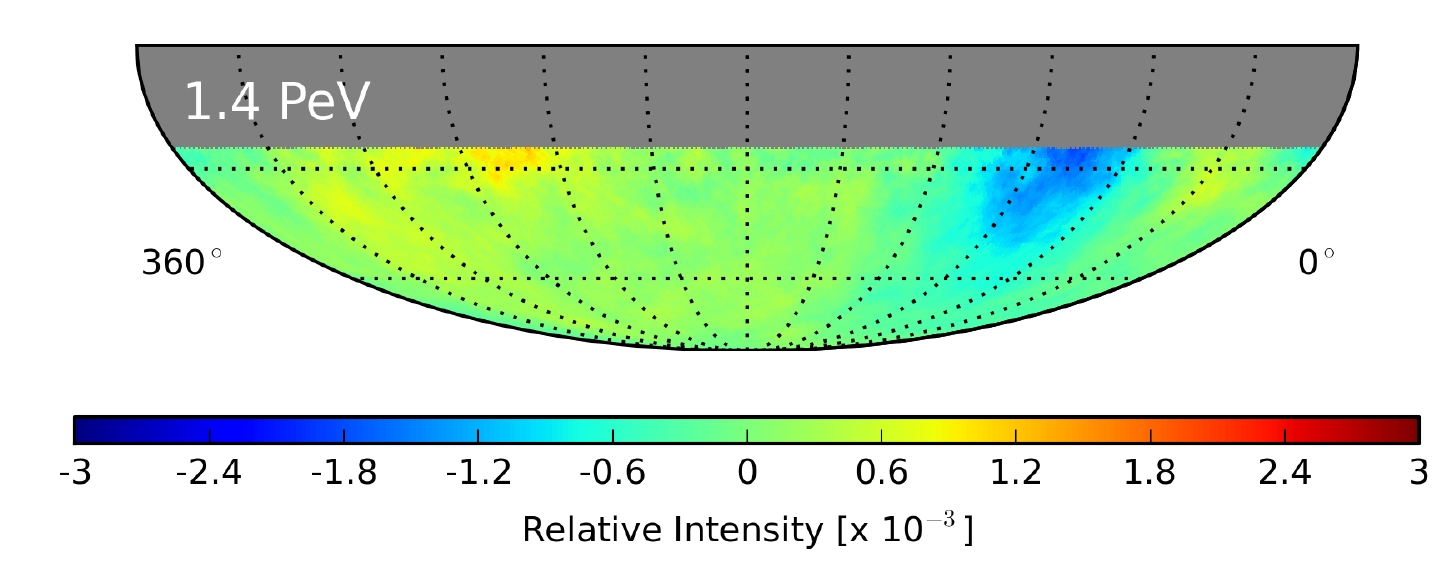}
  \includegraphics[width=0.49\textwidth]{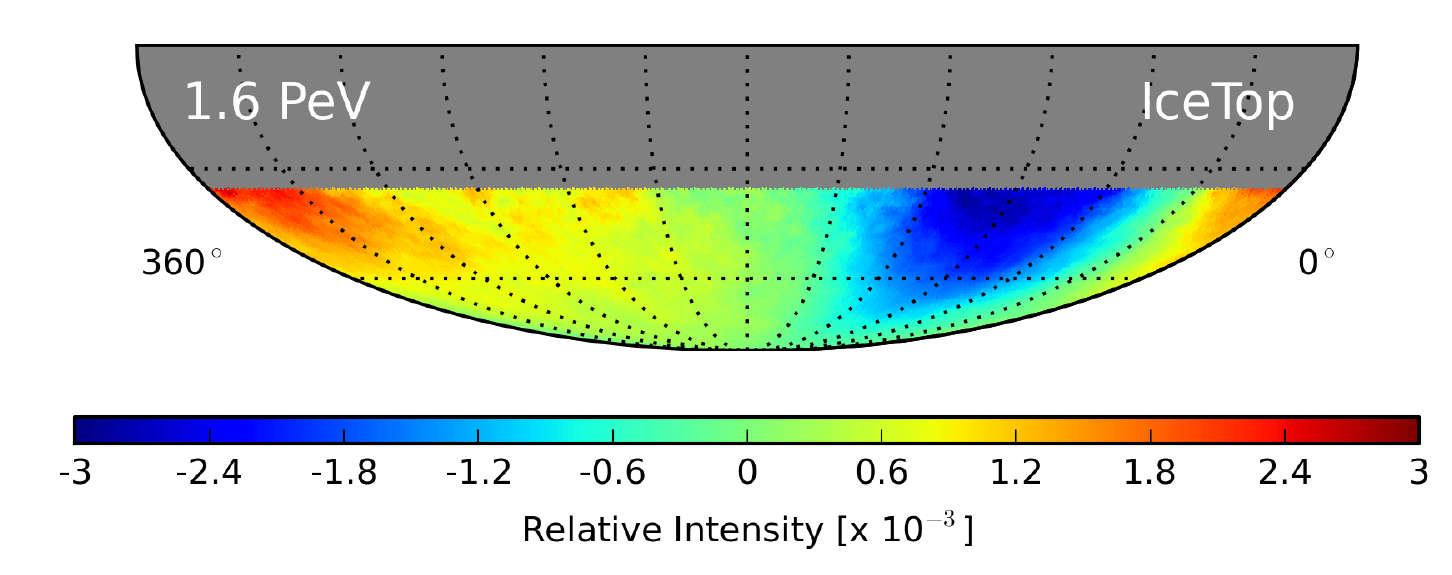}
  \includegraphics[width=0.49\textwidth]{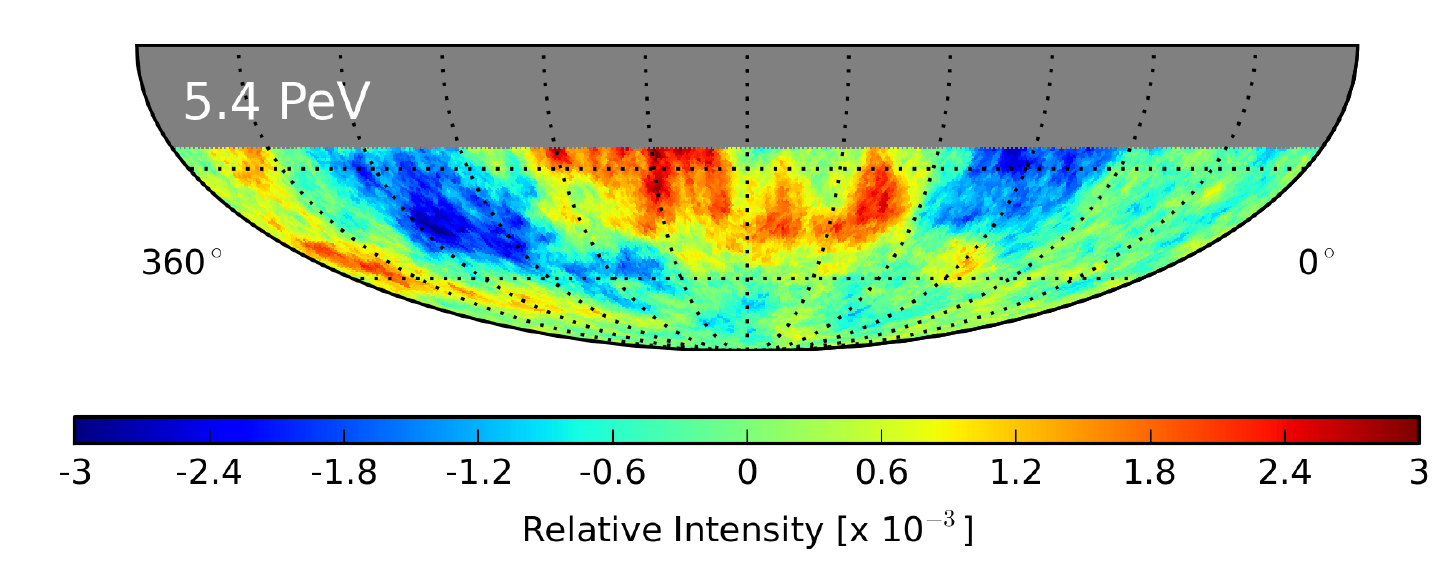}
  \caption{Relative intensity maps in equatorial coordinates for the energy
bins described in Section\,\ref{subsec:energy}.  The median energy of the data shown 
in each map is indicated in the upper left.  Maps have been smoothed with a $20^{\circ}$ 
smoothing radius.  The final three maps are shown on a different relative intensity scale.  
The map at 1.6\,PeV in the lower left panel is based on IceTop data.  All other maps show 
IceCube data.}
  \label{fig:eplots}
\end{figure*}
\begin{figure*}[ht]
  \centering1
  \includegraphics[width=0.49\textwidth]{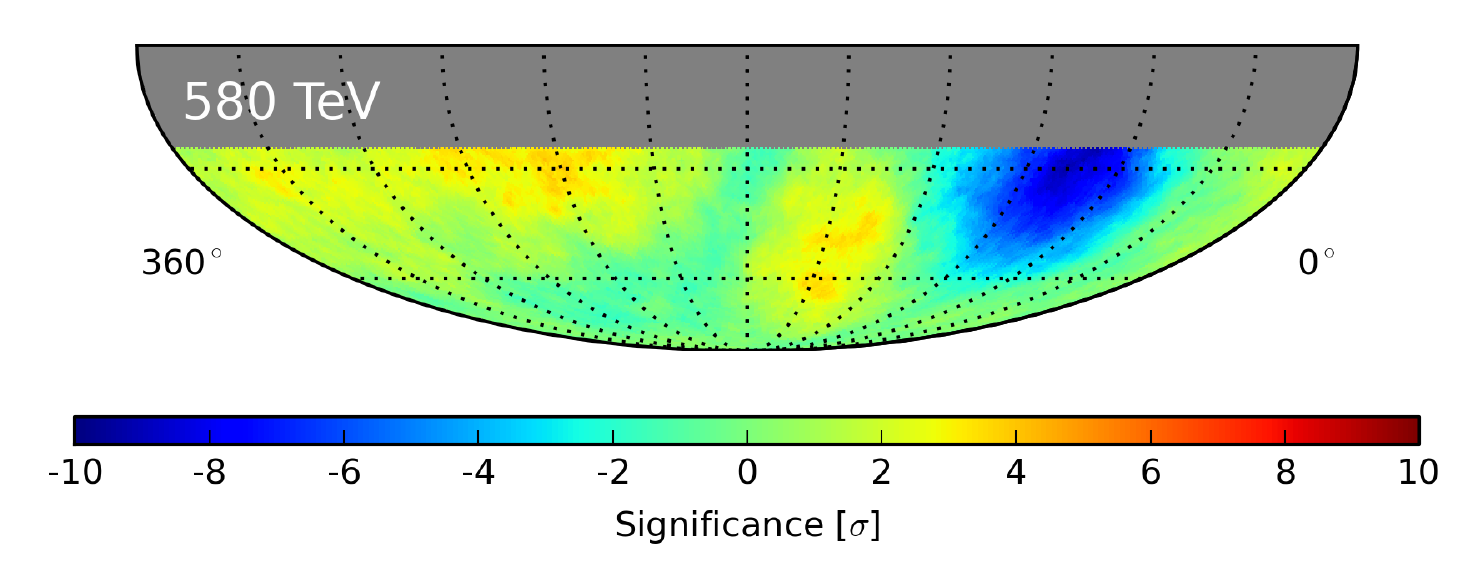}
  \includegraphics[width=0.49\textwidth]{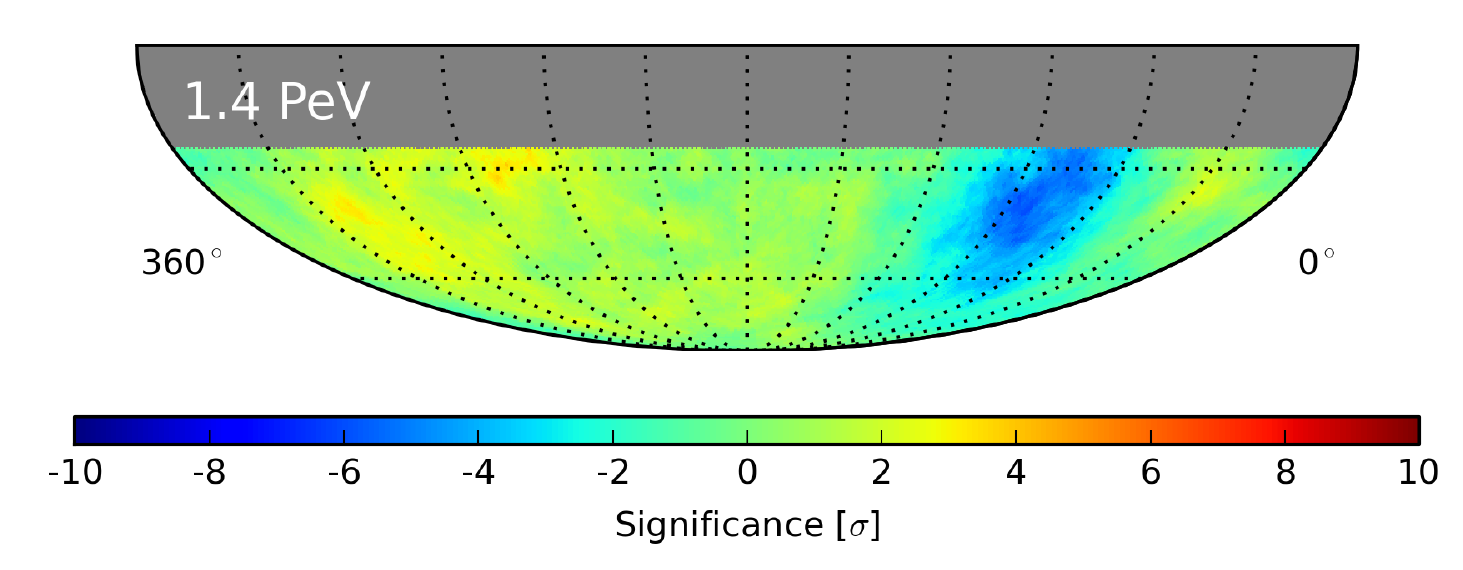}
  \includegraphics[width=0.49\textwidth]{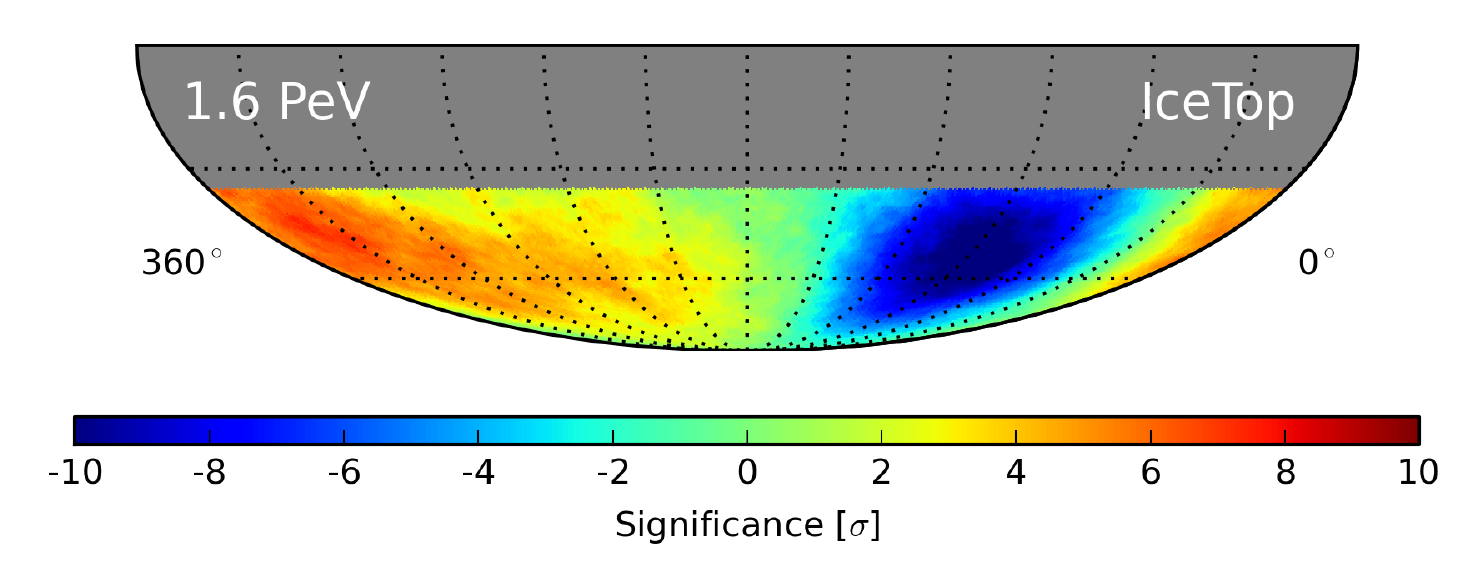}
  \includegraphics[width=0.49\textwidth]{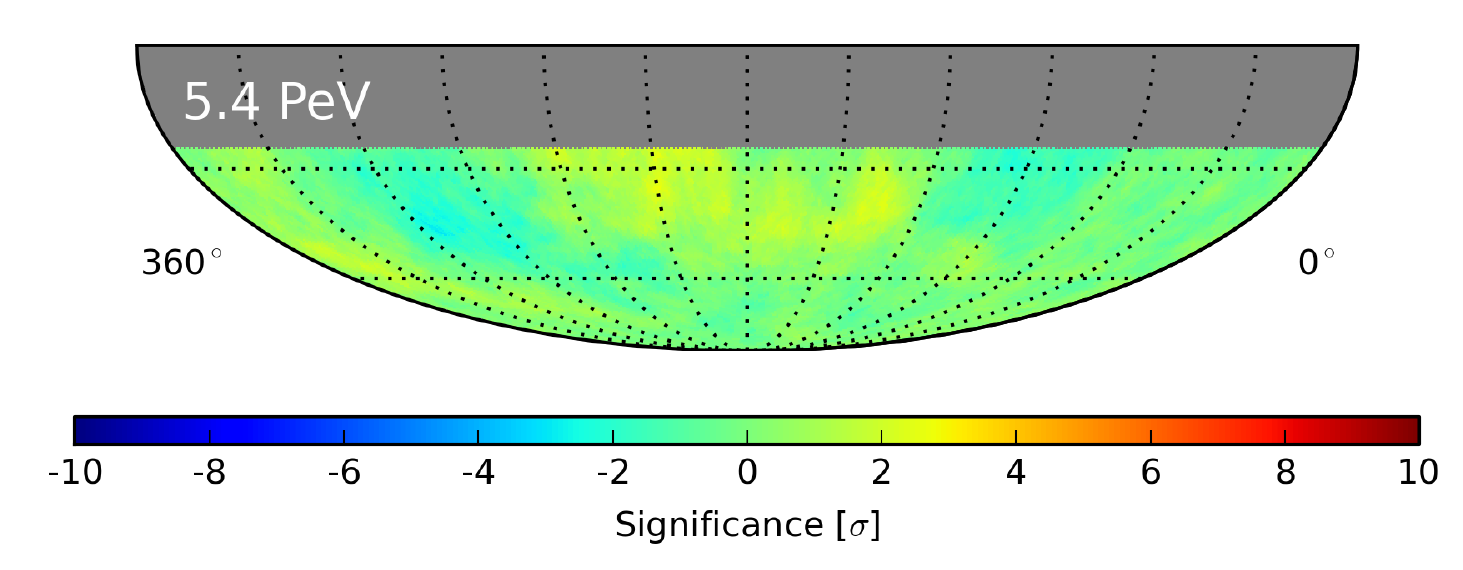}
  \caption{Significance maps in equatorial coordinates for the four highest energy
bins described in Section\,\ref{subsec:energy}.  The median energy of the data shown 
in each map is indicated in the upper left.  Maps have been smoothed with a $20^{\circ}$ 
smoothing radius.  The map at 1.6\,PeV in the lower left panel is based on IceTop data. 
All other maps show IceCube data.}
  \label{fig:splots}
\end{figure*}
\begin{figure*}[p]
\begin{tabular}{ccc}
  \includegraphics[width=0.32\textwidth]{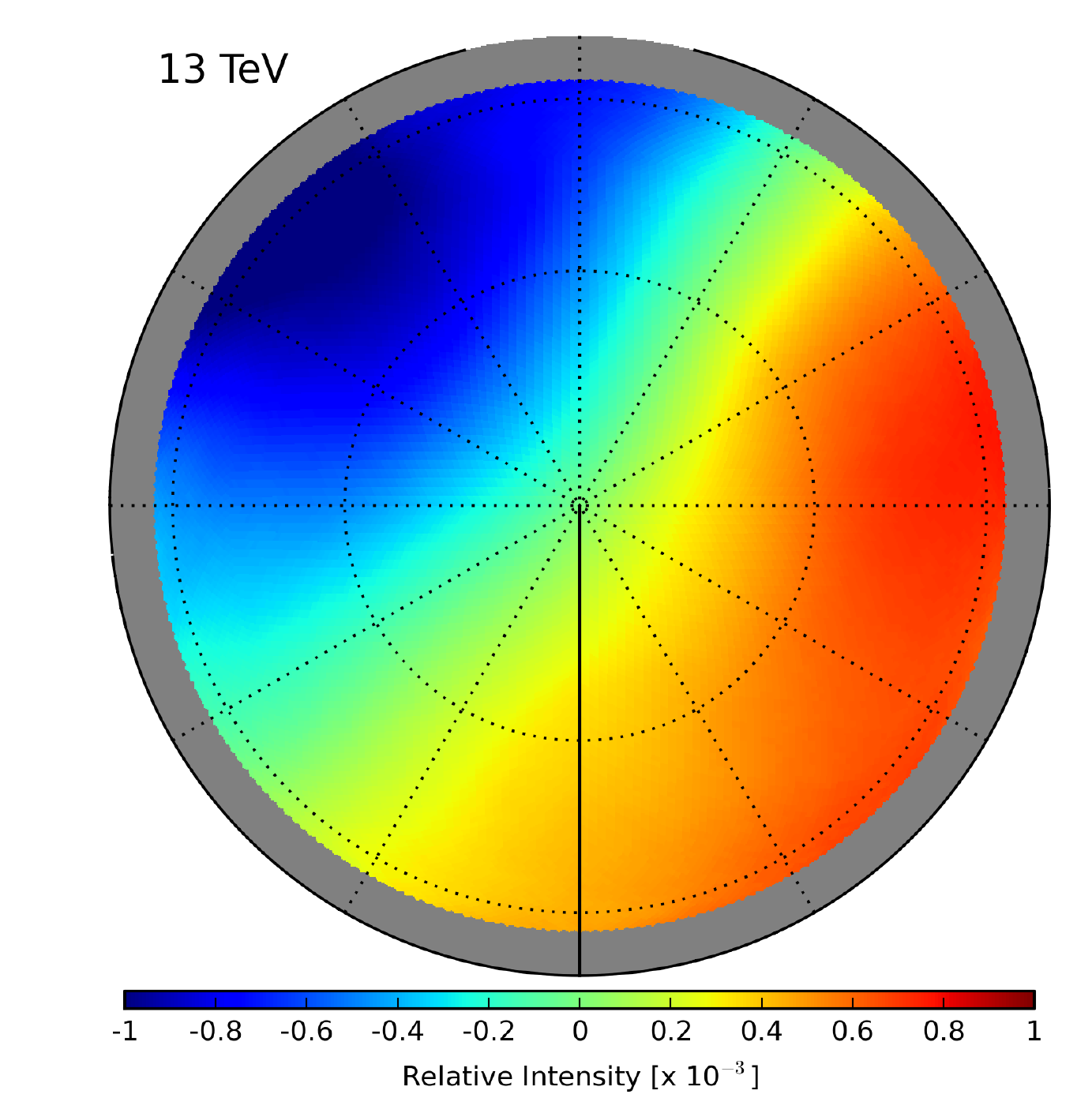} &
\hskip-.5cm
  \includegraphics[width=0.32\textwidth]{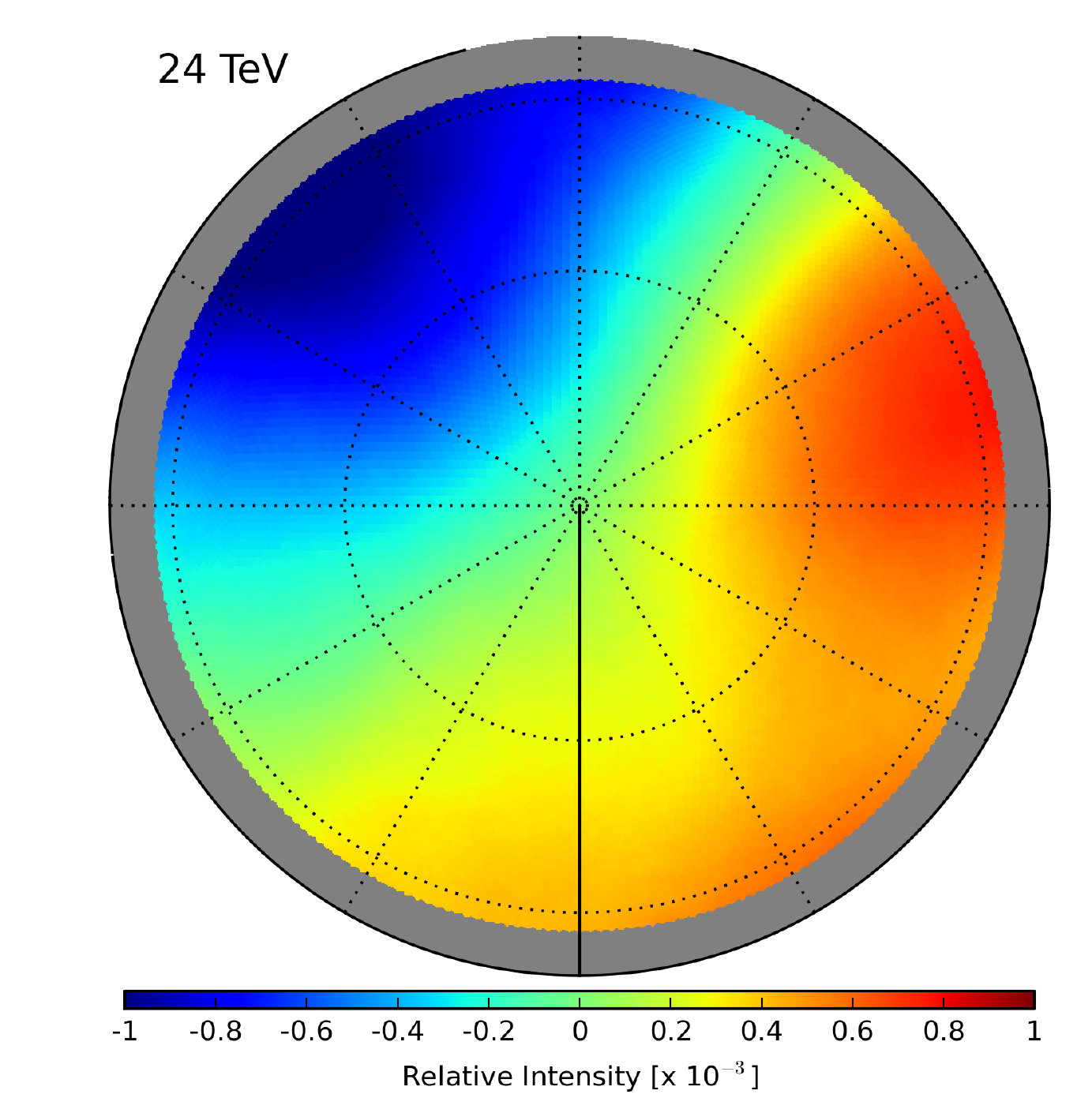} &
\hskip-.5cm
  \includegraphics[width=0.32\textwidth]{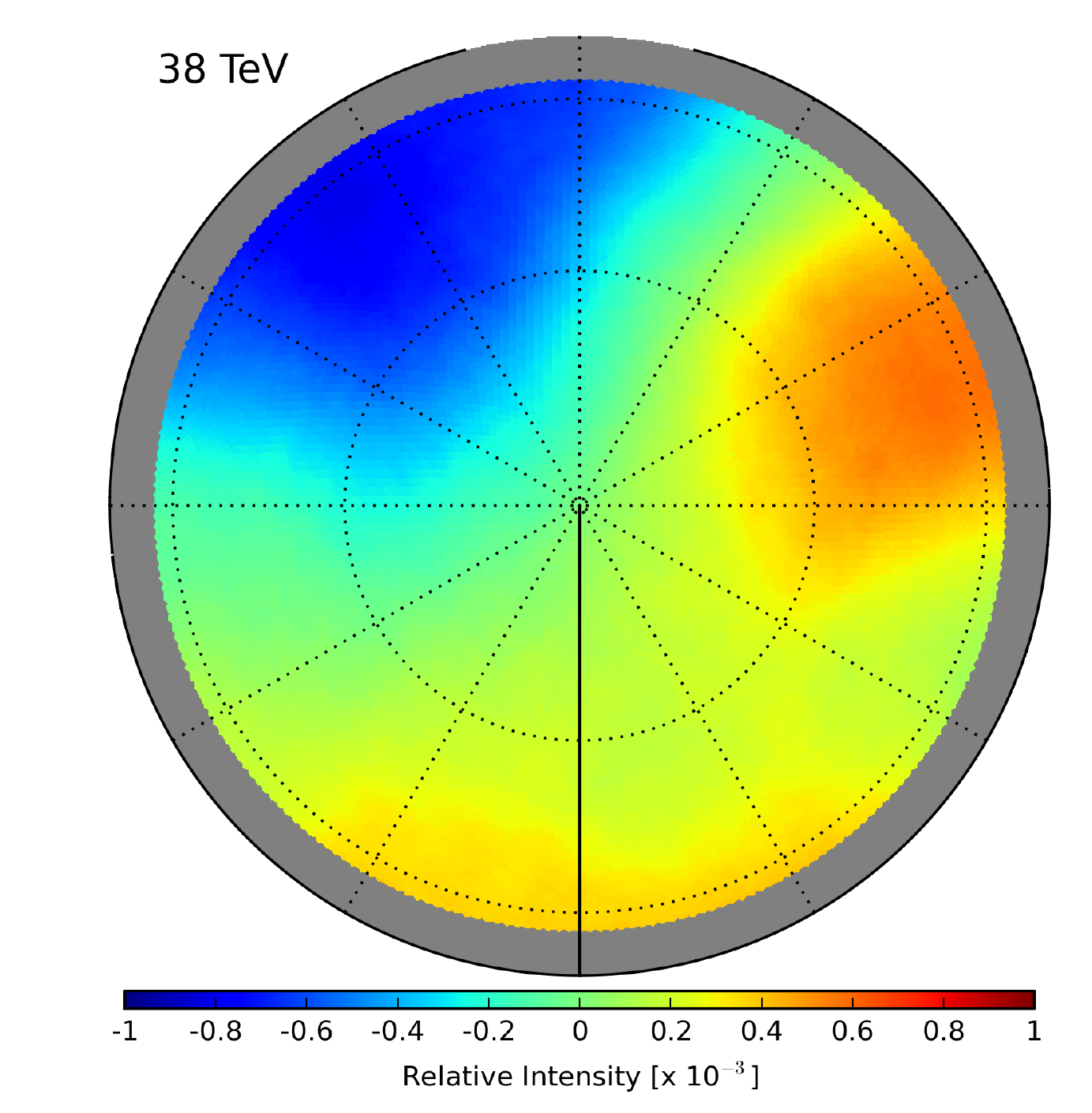}\\
  \includegraphics[width=0.32\textwidth]{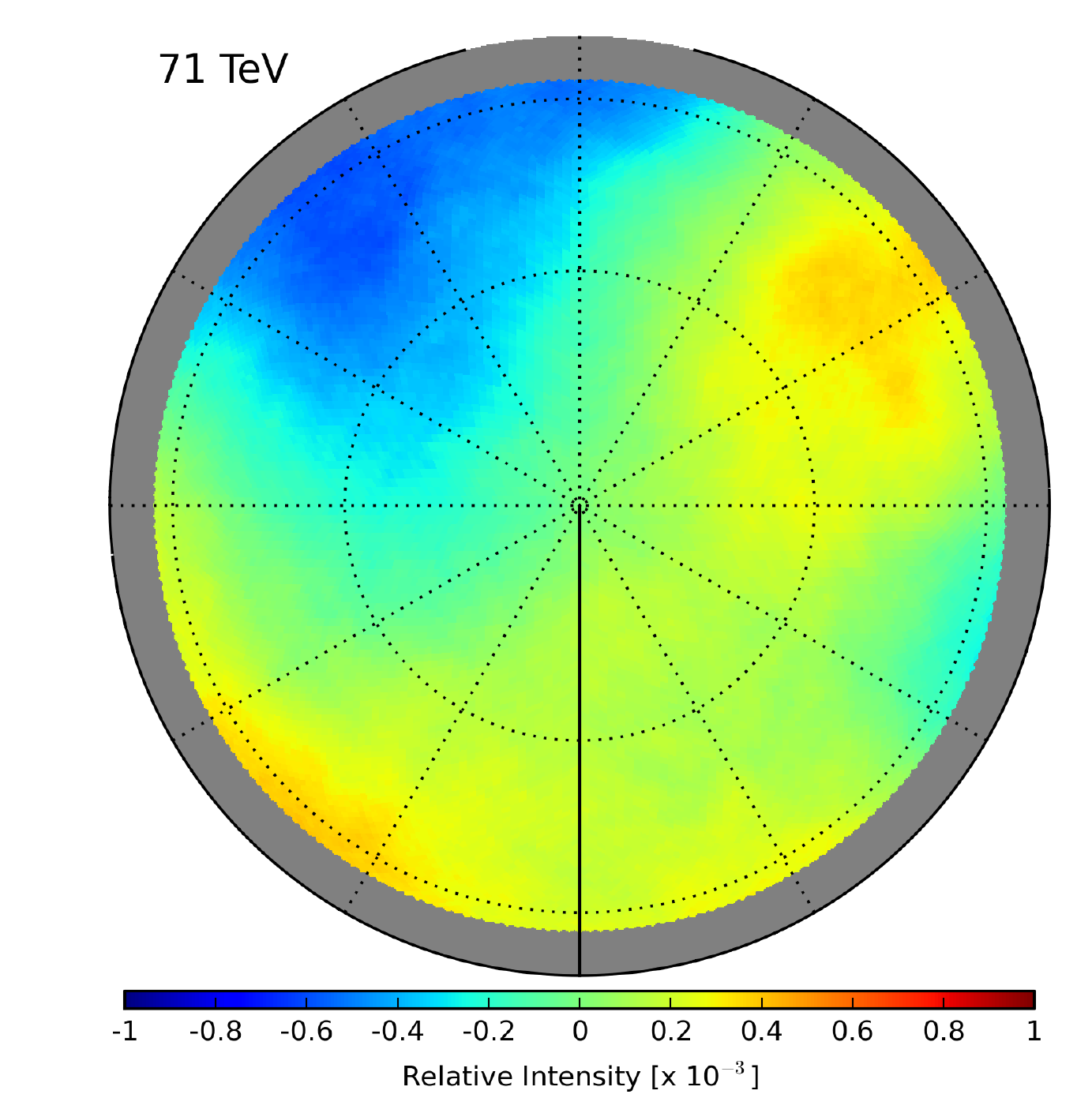} &
\hskip-.5cm
  \includegraphics[width=0.32\textwidth]{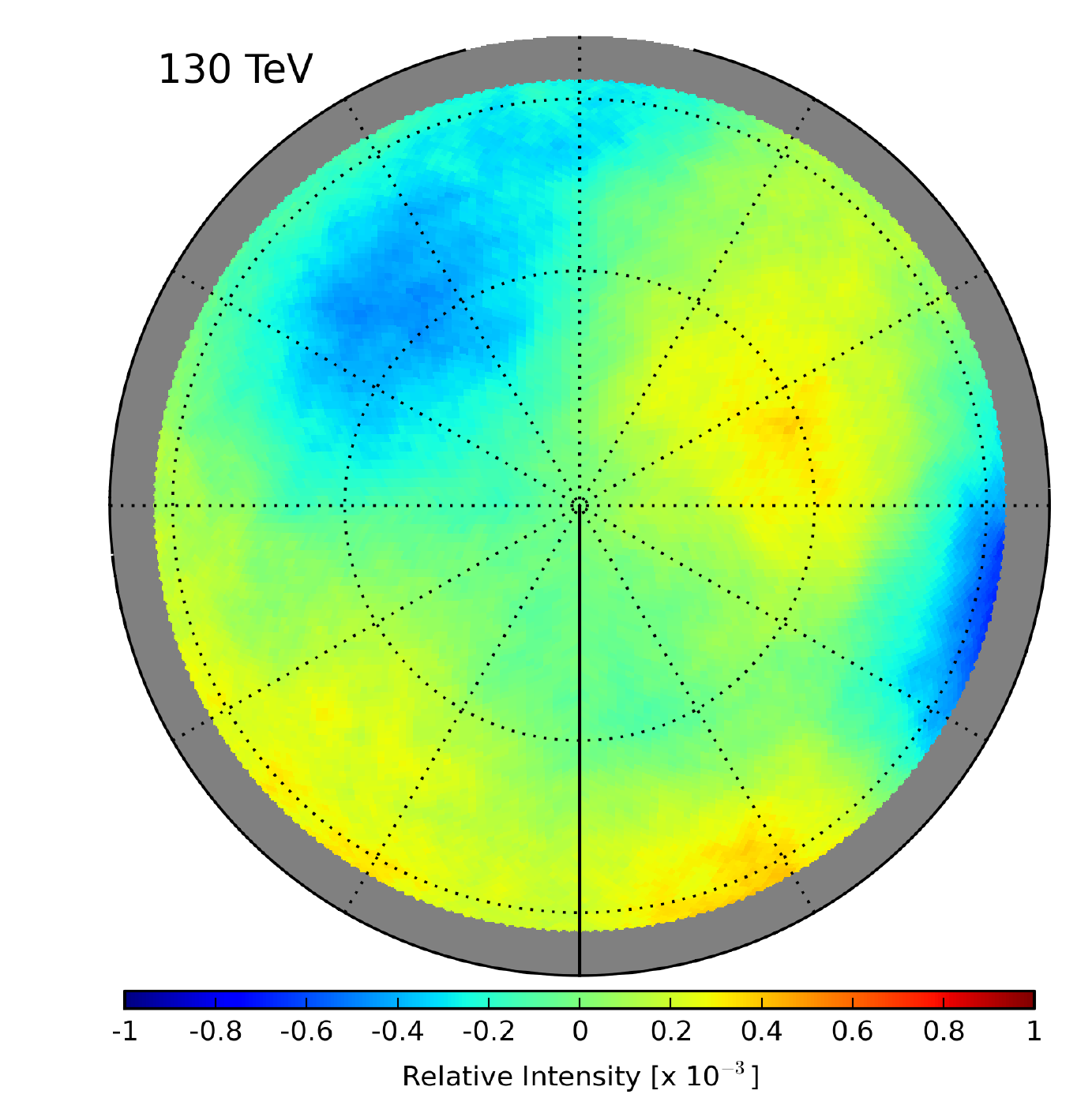} &
\hskip-.5cm
  \includegraphics[width=0.32\textwidth]{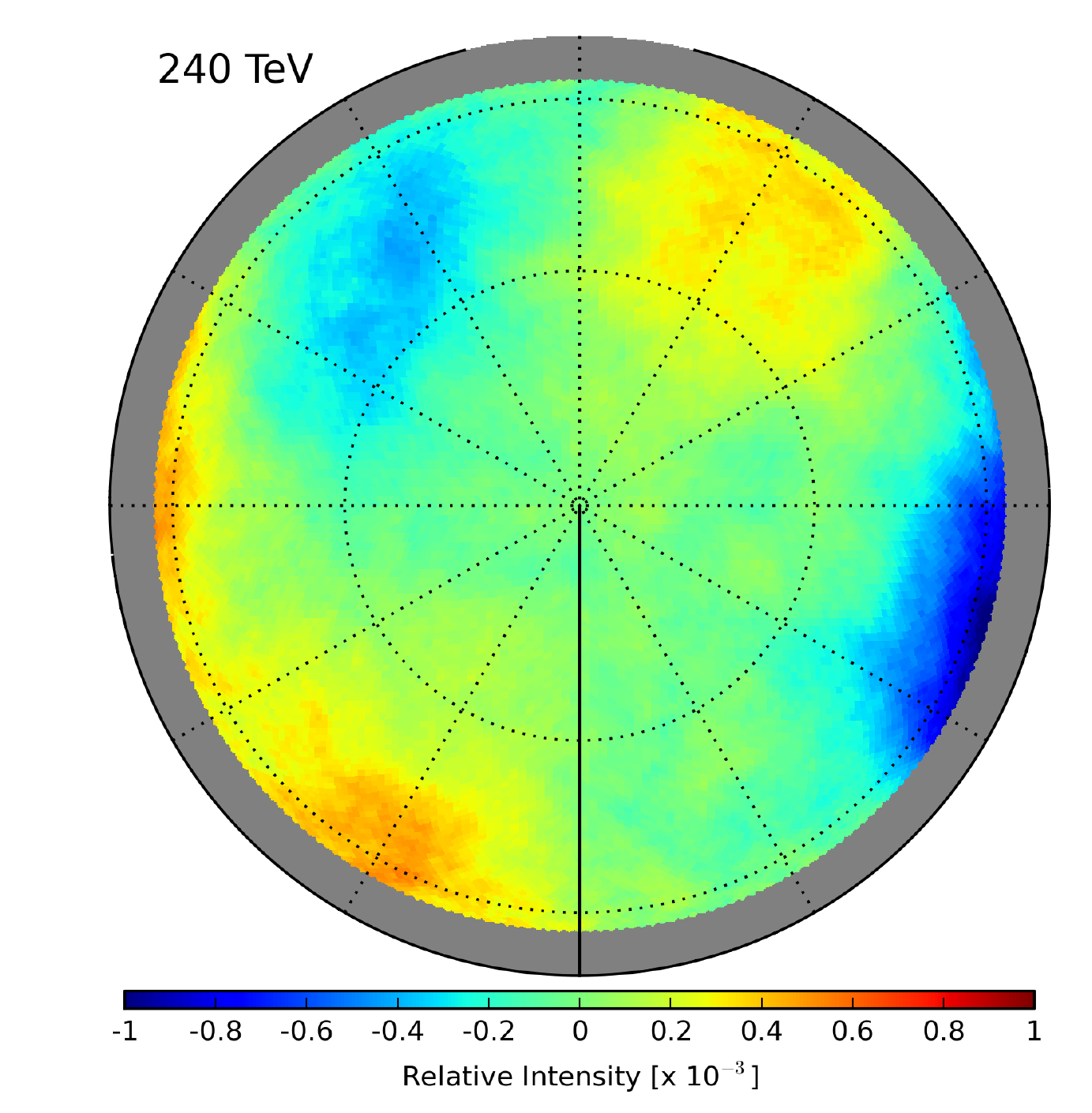}\\
  \includegraphics[width=0.32\textwidth]{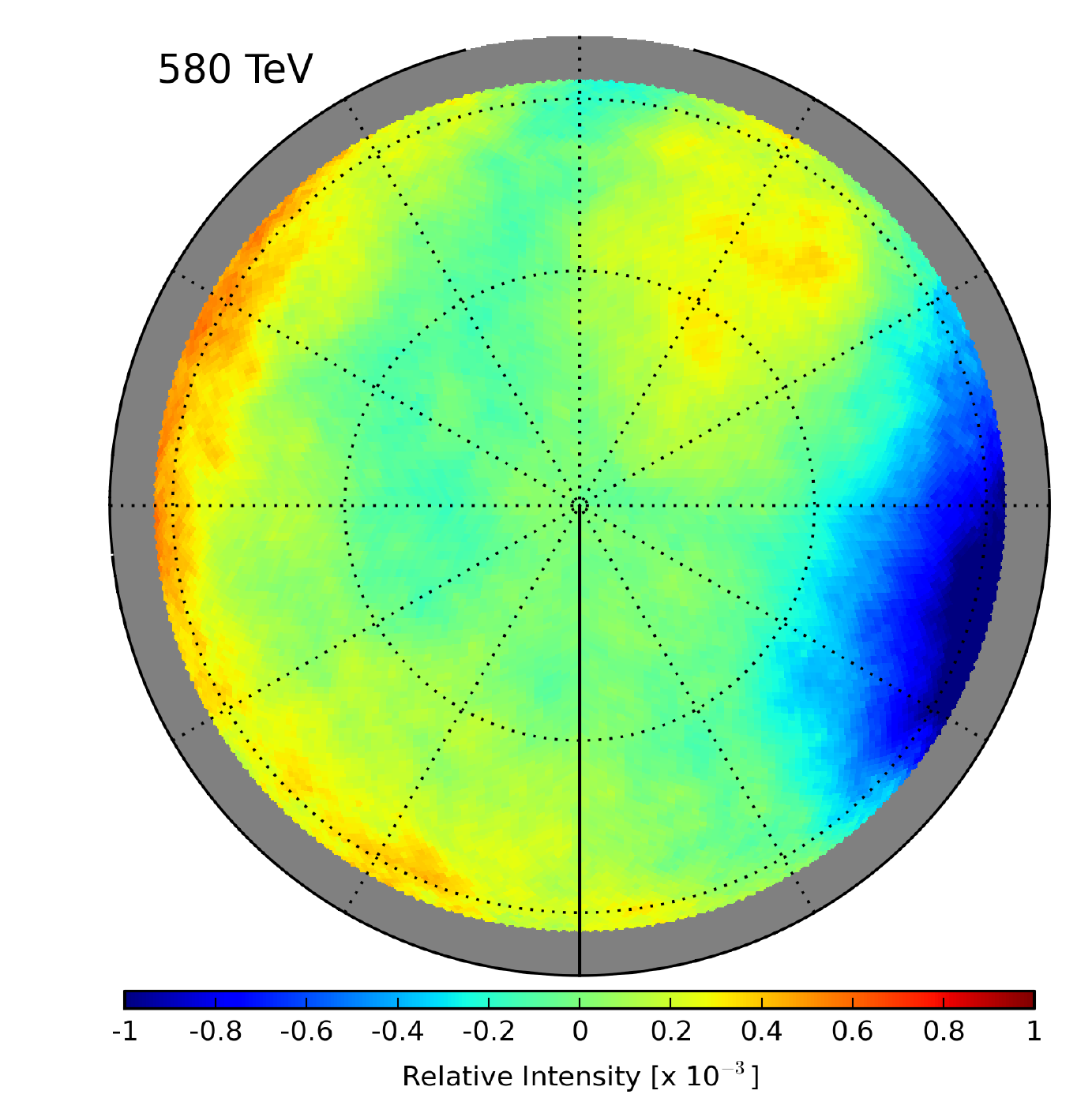} &
\hskip-.5cm
  \includegraphics[width=0.32\textwidth]{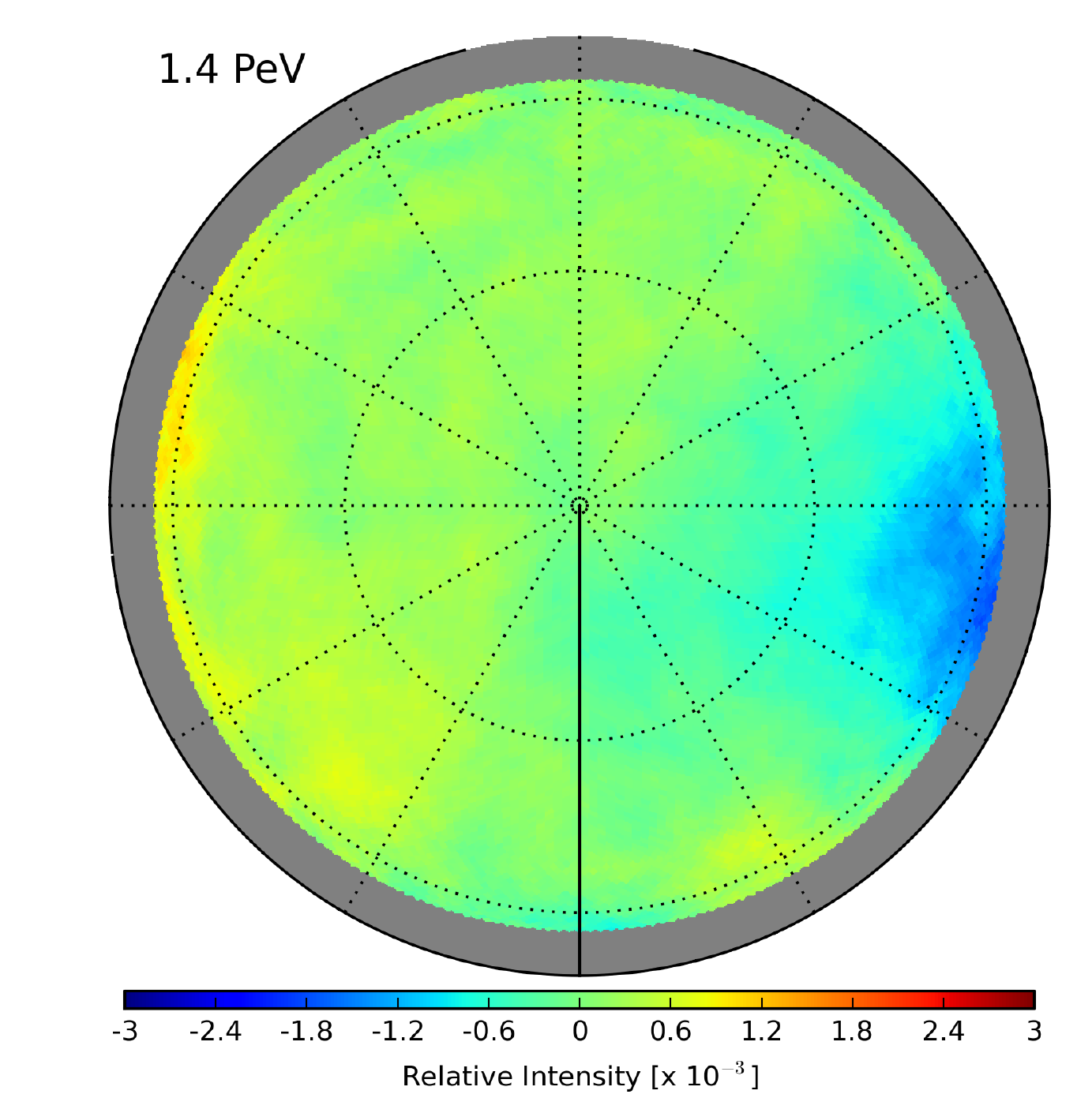} &
\hskip-.5cm
  \includegraphics[width=0.32\textwidth]{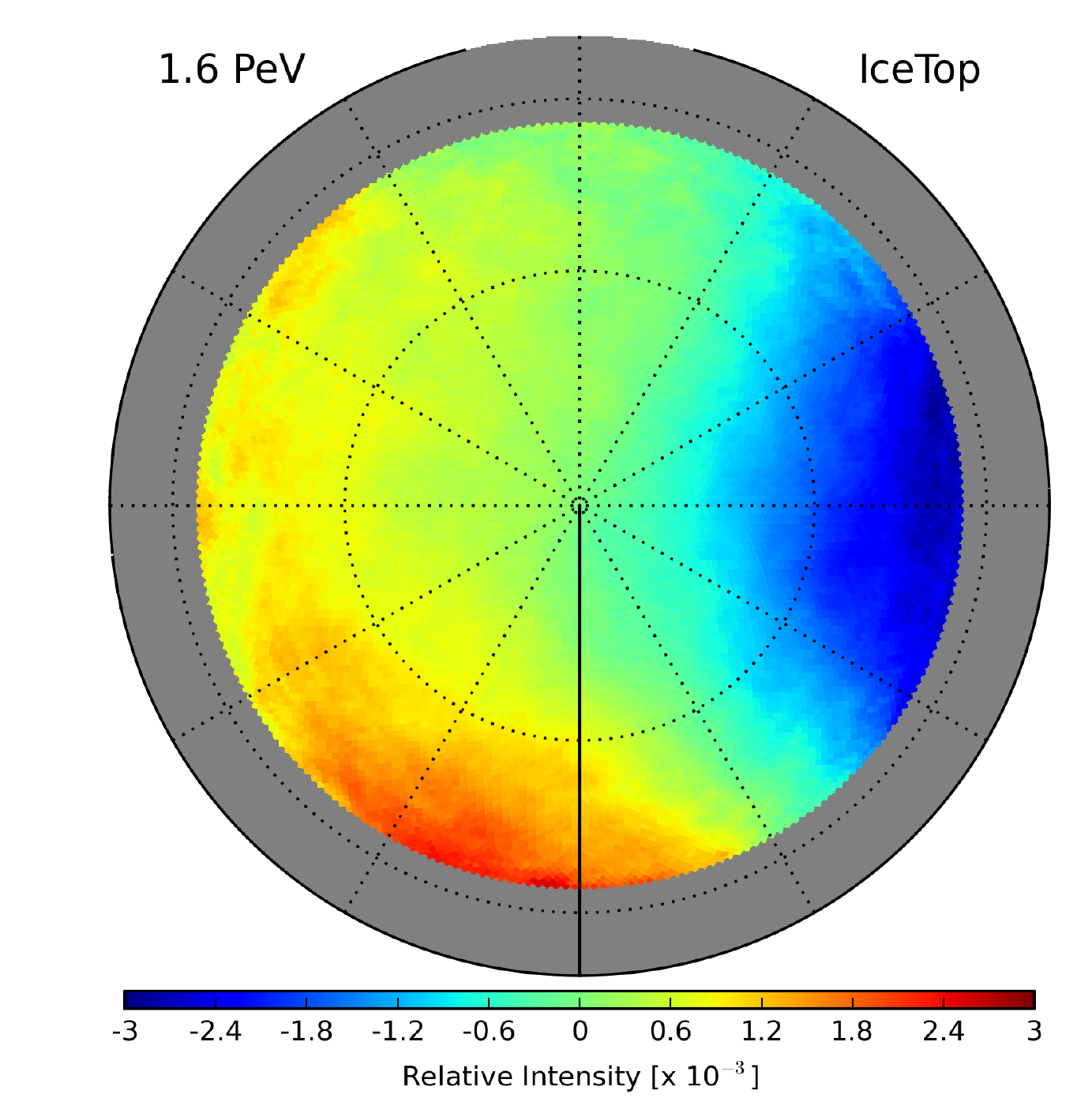}\\
  & 
\hskip-.5cm
  \includegraphics[width=0.32\textwidth]{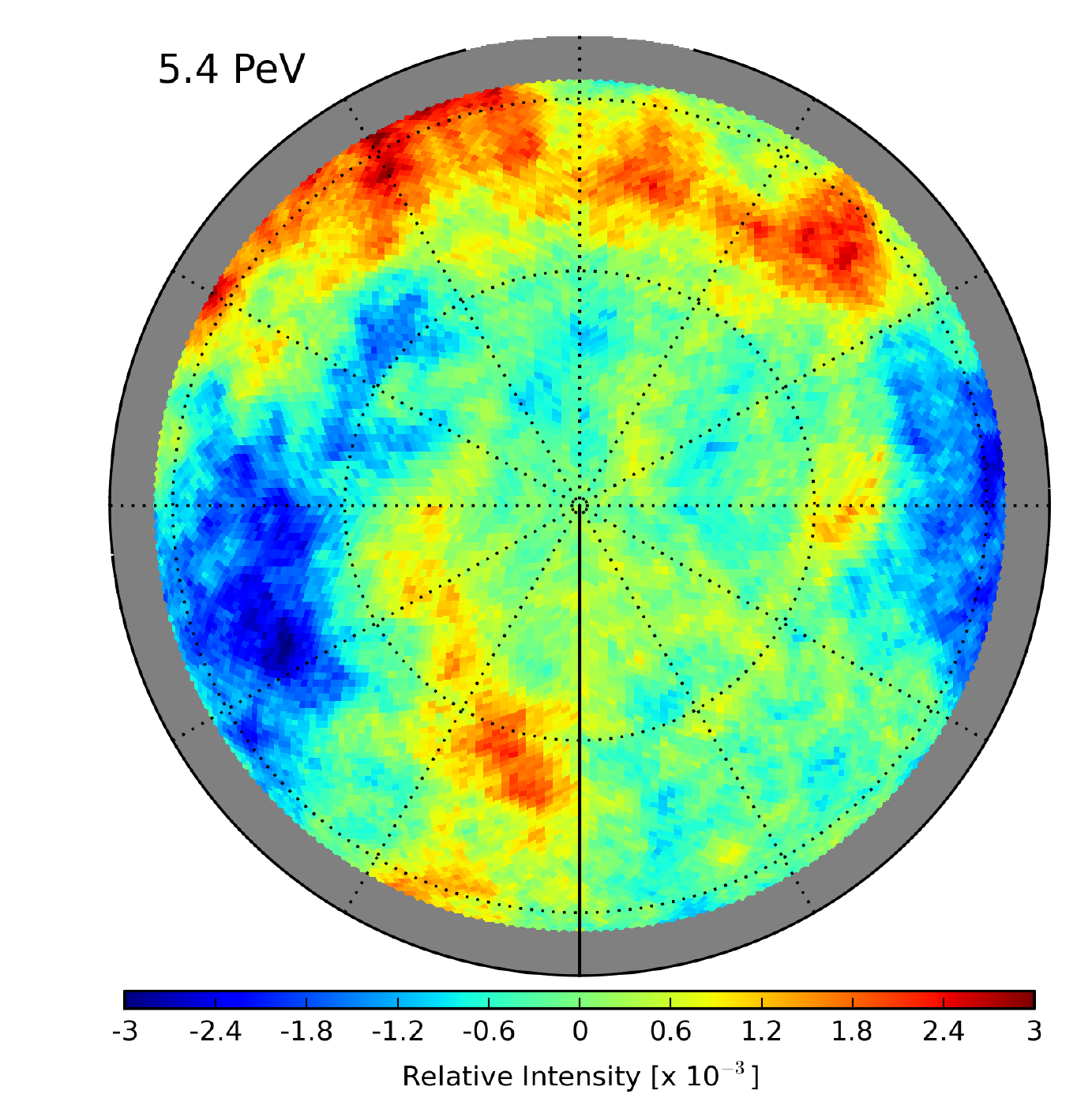} &
\end{tabular}
\vskip-0.3cm
  \caption{Relative intensity maps in polar coordinates for the energy
bins described in Section\,\ref{subsec:energy}.  The median energy of the data shown 
in each map is indicated in the upper left.  Maps have been smoothed with a $20^{\circ}$ 
smoothing radius.  The final three maps are shown on a different relative intensity scale.  
The 1.6\,PeV map is based on IceTop data.  All other maps show IceCube data.}
  \label{fig:polar}
\end{figure*}

The angular power spectrum for the six-year data set is shown in Fig.\,\ref{fig:powspec}.
Similar to previous work, it is calculated using PolSpice~\citep{Szapudi:2001feb,
Chon:2004may}, which corrects for systematic effects introduced by partial-sky
coverage.  The power spectrum is calculated for the unsmoothed data map and is shown 
before (\textit{blue}) and after (\textit{red}) subtracting the dipole and quadrupole 
functions from the sky map.  The gray bands indicate the 68\% and 95\% spread in the 
$C_{\ell}$ for a large number of power spectra for isotropic data sets generated by 
introducing Poisson fluctuations in the reference skymap.  The power spectrum confirms 
the presence of significant structure up to multipoles $\ell\simeq20$, corresponding 
to angular scales of less than $10^{\circ}$.

The error bars on the $C_{\ell}$ shown in Fig.\,\ref{fig:powspec} are statistical.  
We estimate the systematic error caused by the partial-sky coverage by comparing the 
angular power spectrum before and after subtraction of the best-fit dipole and quadrupole 
functions.  After the subtraction, $C_{1}$ and $C_{2}$ are consistent with zero, as
expected.  In principle, the two spectra should be identical for all $\ell\geq3$, but 
because of the partial-sky coverage, the multipole moments are no longer independent. 
While PolSpice tries to mitigate the effect of coupling between multipole moments, a 
significant coupling between the low-$\ell$ modes remains.  As a consequence, the
subtraction of dipole and quadrupole fits also leads to a strong reduction in the 
power of the $\ell=3$, $\ell=4$, and $\ell=5$ multipoles.  The systematic error on 
these multipoles is therefore large, as we cannot rule out that the presence of these 
multipoles is entirely caused by systematic effects.  For multipoles $\ell\geq 6$, 
the distortion is much smaller and the spectra agree within uncertainties.  For these 
moments, the systematic errors on the $C_{\ell}$ are therefore at most of the same 
order as the statistical errors.

In the unsubtracted power spectrum, the uncertainty in the lower multipole 
moments causes the $C_{\ell}$ value for $\ell=5$ to be negative --- a result of
PolSpice's calculation of the $C_{\ell}$ values through the use of the
two-point autocorrelation function.  Simulations using artificial sky maps 
with strong dipole components indicate that this behavior is typical for the
weighting and apodization used in this analysis (see~\citet{IceCube:2011oct} for 
details) and is another indication of the coupling between low-$\ell$ multipoles.

\subsection{Energy Dependence of Anisotropy}
\label{subsec:edependence}

To study the energy dependence of the cosmic-ray anisotropy, we split the data 
into the nine energy bins described in Section\,\ref{subsec:energy}.  This results
in a sequence of maps with increasing median energy, starting from 13\,TeV for the
lowest-energy bin to 5.3\,PeV for the highest-energy bin.  The sky maps in relative 
intensity for all nine energy bins in equatorial coordinates are shown in
Fig.\,\ref{fig:eplots}.  In addition to the nine maps based on IceCube data, we also 
show the IceTop map with its median energy of 1.6\,PeV.  Because of the reduced 
statistics in these maps, we have applied a top-hat smoothing procedure with a 
smoothing radius of $20^{\circ}$ to all, improving the sensitivity to larger structure.
Note that the relative intensity scale for these plots is identical for energies 
up to 580\,TeV, where it then switches to a different scale to account for the strong
increase in relative intensity.  For the IceTop bins with 580\,TeV, 1.4\,PeV, and 5.4\,PeV
median energy and for the IceTop data, Fig.\,\ref{fig:splots} shows the sky maps
in statistical significance.

The maps clearly indicate a strong energy dependence of the global anisotropy. 
The large excess from $30^{\circ}$ to $120^{\circ}$ and deficit from $150^{\circ}$ 
to $250^{\circ}$ that dominate the sky map at lower energies gradually disappear 
above 50\,TeV.  Above 100 TeV a change in the morphology is observed.  At higher
energies, the anisotropy is characterized by a wide relative deficit from $30^{\circ}$ 
to $120^{\circ}$, with an amplitude increasing with energy up to at least 5\,PeV, the 
highest energies currently accessible to IceCube.  To illustrate the phase change,
the relative intensity sky maps are shown in polar coordinates in Fig.\,\ref{fig:polar}.
It is important to note that the time-scrambling method used to calculate the 
reference map decreases in sensitivity as we approach the polar regions.  This effect
is clearly visible in Fig.\,\ref{fig:polar}, where the relative intensity approaches 
zero at the pole for each map, but is not indicative of the morphology of the true 
anisotropy.

Because of the poor energy resolution, it is difficult to accurately determine 
the energy where the transition in anisotropy occurs and how rapid the transition is.
To illustrate the energy dependence of the phase and strength of the anisotropy, we 
show in Fig.\,\ref{fig:dipole} amplitude (\textit{left}) and phase (\textit{right}) 
of the dipole moment as a function of energy.  Both values are calculated by fitting 
the set of harmonic functions with $n\leq 3$ to the projection of the two-dimensional 
relative intensity map (Fig.\,\ref{fig:eplots}) in right ascension,
\begin{equation}
\sum_{n=0}^{3} A_{n} \cos[n(\alpha-\phi_{n})]~, 
\label{eq1}
\end{equation}
where $A_{n}$ is the amplitude and $\phi_{n}$ is the phase of the $n^{th}$ harmonic 
term, respectively.  The fit is performed on a projection with a $5^{\circ}$ bin width 
in right ascension.  We fit the one-dimensional projection in right ascension
rather than the full sky map because the two-dimensional fit of spherical harmonics to 
the map is difficult to perform with a limited field of view.  As a result of the method 
we apply to generate the reference map, the sky map will in any case only show the 
projection of any dipole component, so the one-dimensional fit is sufficient to study 
the energy dependence of the dominant dipole. The values for the projections in each 
energy bin are provided in Tab.\,\ref{tab:proj}.

\begin{table*}[t]
\resizebox{\textwidth}{!}{\begin{tabular}{ r | r r r r r r r r r | r}
    \hline
     R.A. &
\multicolumn{10}{c}{$\log_{10}(E_{\mathrm{median}}/\mathrm{GeV})$}\\
     & $4.12^{+0.62}_{-0.50}$ & $4.38^{+0.65}_{-0.54}$ & $4.58^{+0.68}_{-0.55}$
     & $4.85^{+0.73}_{-0.64}$ & $5.12^{+0.74}_{-0.72}$ & $5.38^{+0.75}_{-0.78}$
     & $5.77^{+0.60}_{-0.83}$ & $6.13^{+0.52}_{-0.63}$ & $6.73^{+0.46}_{-0.58}$
     & $6.21^{+0.36}_{-0.27}$ \\
    \hline
     \multicolumn{1}{l|}{$350^{\circ}$} & 3.84 &   3.56 &  3.11 &  1.41 &  0.89 &  1.29 &  0.36 &  -3.00 &   6.64 &  18.29 \\
     \multicolumn{1}{l|}{$330^{\circ}$} & 2.64 &   3.45 &  3.41 &  2.96 &  1.81 &  4.36 &  1.72 &   0.90 &  23.10 &  15.79 \\
     \multicolumn{1}{l|}{$310^{\circ}$} & 0.34 &   1.17 &  0.78 &  1.52 &  2.32 &  2.77 &  1.46 &   8.89 &  -2.14 &   8.99 \\
     \multicolumn{1}{l|}{$290^{\circ}$} & -3.02 &  -1.52 & -0.16 &  0.19 &  1.57 &  1.26 &  2.22 &   1.60 & -40.43 &   6.56 \\
     \multicolumn{1}{l|}{$270^{\circ}$} & -4.46 &  -3.25 & -1.72 & -0.76 &  0.64 &  1.11 &  1.14 &   6.56 &  -2.97 &  10.44 \\
     \multicolumn{1}{l|}{$250^{\circ}$} & -7.11 &  -6.33 & -3.77 & -2.63 & -2.15 & -0.46 &  2.87 &   9.04 &  -8.79 &   6.02 \\
     \multicolumn{1}{l|}{$230^{\circ}$} & -10.30 & -10.24 & -7.88 & -5.59 & -4.49 & -3.89 &  2.09 &   0.85 &   8.49 &  11.14 \\
     \multicolumn{1}{l|}{$210^{\circ}$} & -9.24 &  -9.02 & -7.24 & -5.51 & -3.13 & -1.71 &  1.54 &   5.26 &  15.01 &   8.23 \\
     \multicolumn{1}{l|}{$190^{\circ}$} & -7.14 &  -6.74 & -5.66 & -2.70 & -3.07 & -0.25 &  0.46 &  -0.25 &   8.10 &  -0.99 \\
     \multicolumn{1}{l|}{$170^{\circ}$} & -5.26 &  -5.26 & -4.27 & -3.78 & -1.52 &  2.07 & -1.21 &   0.75 &  -7.95 &   2.75 \\
     \multicolumn{1}{l|}{$150^{\circ}$} & 0.05 &   0.32 &  1.54 &  3.15 &  2.16 &  3.93 &  5.81 &   1.47 &  39.23 &  -1.61 \\
     \multicolumn{1}{l|}{$130^{\circ}$} & 4.03 &   3.83 &  2.82 &  2.08 &  1.47 &  1.07 & -1.06 &  -0.89 &  -2.84 & -11.94 \\
     \multicolumn{1}{l|}{$110^{\circ}$} & 6.66 &   7.41 &  6.62 &  3.92 &  2.38 & -1.70 & -1.46 &  -4.50 &  -5.75 & -21.35 \\
     \multicolumn{1}{l|}{$90^{\circ}$} & 6.92 &   6.09 &  3.84 &  1.95 &  0.84 & -0.71 & -5.38 & -11.09 & -11.34 & -21.95 \\
     \multicolumn{1}{l|}{$70^{\circ}$} & 6.06 &   4.52 &  1.39 & -0.16 & -1.53 & -7.26 & -8.70 & -14.24 &  10.69 & -20.22 \\
     \multicolumn{1}{l|}{$50^{\circ}$} & 5.89 &   4.18 &  2.89 &  0.31 & -1.32 & -0.74 & -2.51 &  -1.29 &  -3.25 & -17.69 \\
     \multicolumn{1}{l|}{$30^{\circ}$} & 5.37 &   4.35 &  2.23 &  2.21 &  3.12 & -0.33 & -0.05 &  -0.13 & -21.37 &   0.00 \\
     \multicolumn{1}{l|}{$10^{\circ}$} & 4.80 &   3.42 &  2.08 &  1.55 & -0.14 & -0.74 &  1.92 &  -1.19 &  11.77 & 12.26 \\
    \hline
     \multicolumn{1}{l|}{$\sigma_{\mathrm{stat}}$} & 0.18 & 0.18 & 0.36 & 0.66 & 1.02 & 1.42 & 1.49 & 3.23 & 14.39 & 4.48 \\
    \hline
  \end{tabular}}
\caption{Relative intensity ($\times 10^{4}$) projected in right ascension for each energy bin. 
Right ascension values indicate the center of each bin.  Energy values indicate the median true 
energy for each bin, as estimated by simulation, with error bars containing 68\% of the data. 
The last row gives the mean of the statistical error on the relative intensity values in 
the column ($\sigma_{stat}$).  The separate rightmost column contains data for the IceTop energy bin.}
  \label{tab:proj}
\end{table*}
The red data points in Fig.\,\ref{fig:dipole} are based on the IceTop data.  While
the phase agrees well with that of the IceCube data at similar energies, the amplitude 
of the anisotropy is larger for the IceTop data than for any IceCube energy bin.  A 
possible explanation for the difference could be the different chemical composition of 
the IceCube and IceTop data sets.  Table~\ref{tab:comp} shows the relative composition 
of cosmic rays detected in IceCube and IceTop according to simulation, based on a primary 
cosmic-ray composition according to the model by~\citet{Hoerandel:2002yg}.  For IceCube, 
we list the composition for all nine energy bins.   Elements are grouped in four main 
categories with increasing mass number as described in the caption.  The simulation 
indicates that the data set recorded by IceTop is composed of 34\% protons and 12\% heavy 
elements.  At a comparable median energy, in the second-highest energy bin, the data set 
recorded by IceCube is composed of 24\% protons and 21\% heavy elements.  The reason for 
the discrepancy is the fact that at this median energy, the effective area of IceTop for 
iron showers is still smaller than for proton showers.  Iron primaries start interacting 
higher in the atmosphere than proton primaries, so iron and proton showers are at different 
stages of development when reaching the detector altitude.  The probability to reach the 
detector altitude and trigger at low energy is therefore smaller for iron showers than for 
proton showers.  If the anisotropy is predominantly caused by protons, the lighter 
composition of the IceTop data could lead to a stronger dipole amplitude.

The IceCube and IceTop sky maps also show different structures in other parts of the maps, 
but as indicated in Fig.\,\ref{fig:splots}, most of these structures are not statistically
significant, especially near the edge of the field of view.  The large structure with a 
significance of approximately $5\,\sigma$ between $300^{\circ}$ and $360^{\circ}$ in right
ascension and $-30^{\circ}$ and $-60^{\circ}$ in declination in the IceTop sky map is also 
marginally visible in the 1.4 PeV IceCube map, but with a low significance because of the
small size of the data set.
\begin{table}[ht]
  \centering
  \begin{tabular}{lrrrr}
    \hline
    $\log_{10}(E_{\mathrm{median}}/\mathrm{GeV})$ & H & He & CNO & Fe \\
    \hline
    4.12 & 0.74 & 0.21 &  0.04 & 0.01 \\
    4.38 & 0.70 & 0.23 &  0.06 & 0.01 \\
    4.58 & 0.67 & 0.25 &  0.07 & 0.02 \\
    4.85 & 0.61 & 0.27 &  0.09 & 0.03 \\
    5.12 & 0.54 & 0.28 &  0.12 & 0.05 \\
    5.38 & 0.46 & 0.29 &  0.16 & 0.09 \\
    5.77 & 0.35 & 0.30 &  0.21 & 0.14 \\
    6.13 & 0.24 & 0.28 &  0.26 & 0.21 \\
    $6.21^{*}$ & 0.34 & 0.30 &  0.24 & 0.12 \\
    6.73 & 0.17 & 0.18 &  0.28 & 0.37 \\
    \hline
  \end{tabular}
  \caption{Chemical composition of each IceCube energy bin and for IceTop as determined 
by simulation using a primary chemical composition following~\citet{Hoerandel:2002yg}.  
Listed is the relative fraction of the composition group in the detected cosmic-ray flux.  
The composition groups are as follows: 
H: $^{1}\mathrm{H}$, 
He: $^{4}\mathrm{He}$ - $^{11}\mathrm{B}$, 
CNO: $^{12}\mathrm{C}$ - $^{39}\mathrm{K}$,
Fe: $^{40}\mathrm{Ca}$ - $^{56}\mathrm{Fe}$.
The IceTop bin is marked with an asterisk.}
  \label{tab:comp}
\end{table}

\begin{figure*}[!]
  \centering
  \includegraphics[width=0.98\textwidth]{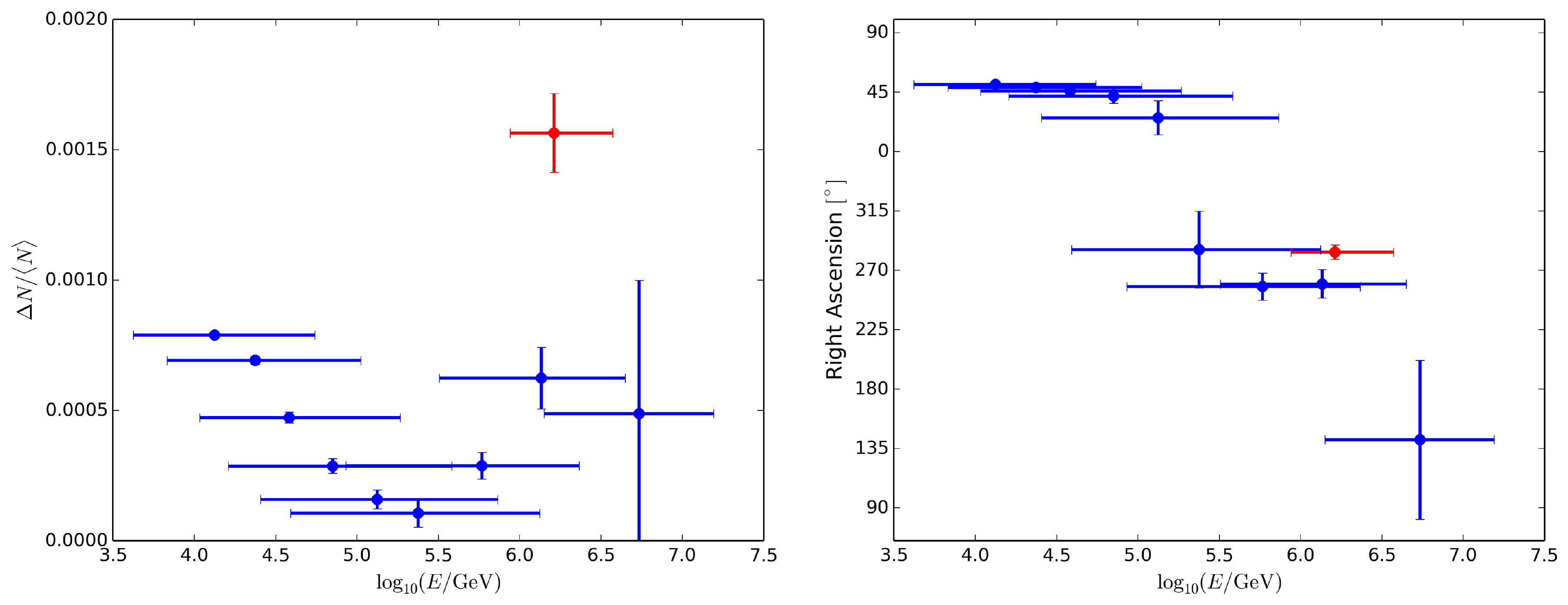}
  \caption{Amplitude ({\it left}) and phase ({\it right}) of the dipole moment of the 
projection of the relative intensity in right ascension for the nine energy bins of 
IceCube ({\it blue}) and for IceTop ({\it red}).  The projections were fit with the 
set of harmonic functions (see Eq.\,\ref{eq1}), but only the dipole is reported here.  
Data points indicate the median energy of each energy bin, with error bars on the energy 
showing the 68\% central 
interval.}
  \label{fig:dipole}
\end{figure*}

\begin{figure*}[ht]
  \centering
    \includegraphics[width=1.0\textwidth]{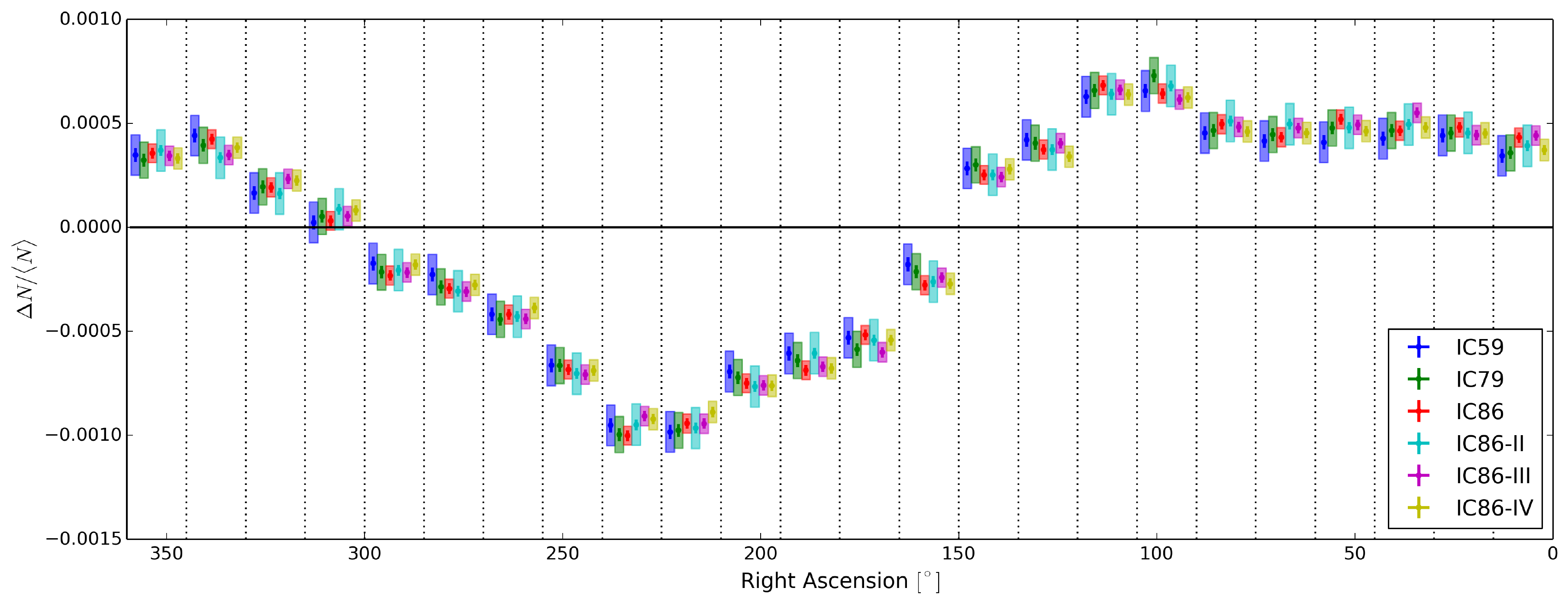}
    \caption{Projection of relative intensity for all declinations as a function of right 
ascension for each configuration of the IceCube detector from IC59 to the fourth year of IC86.
The yearly data points are placed side by side in time sequence, and the different right ascension
bins are delineated by vertical lines.  The shaded areas indicate systematic errors, calculated 
using the anti-sidereal frame for each year independently.}
    \label{fig:proj1dcomp}
\end{figure*}

\subsection{Time Dependence of Anisotropy}
\label{subsec:tdependence}

The data used in this analysis was recorded over a period of six years and therefore
also allows for a study of the stability of the anisotropy over this time period.
An observed time-modulation of the anisotropy, in particular one that coincides with
the 11-year solar cycle, could be evidence for a heliospheric influence on the observations.
Time-dependent studies have been performed previously by several experiments, with
contradictory results.  Milagro reported a steady increase in the amplitude of the 
large-scale anisotropy over a seven-year time period (2000-2007)~\citep{Milagro:2009jun}.
However, the Tibet experiment did not observe significant time variation in the large-scale 
anisotropy between 1999 and 2008~\citep{Tibet:2010mar}, and the ARGO-YBJ experiment did 
not observe significant variation in the medium-scale ($10^{\circ}-45^{\circ}$) anisotropy 
in data covering the period from 2007 to 2012~\citep{ARGO:2013oct}.  Note that the 23rd 
solar cycle lasted from 1996 June to 2008 January and reached a maximum in March 2000.  
The current (24th) solar cycle started in January 2008 and reached a maximum in 
April 2014.  The IceCube data set therefore covers the period from minimum to maximum
of the current cycle.

Figure\,\ref{fig:proj1dcomp} shows the one-dimensional projection of the relative
intensity in right ascension for each detector configuration used in this analysis, 
each one corresponding to approximately a year of data (see Tab.\,\ref{tab:data}).  
The yearly data points are placed side by side in time sequence, and the different
right ascension bins are delineated by vertical lines.  The shaded regions represent 
systematic errors determined by calculating the maximum amplitude of the signal in the 
anti-sidereal time frame (discussed in Section~\ref{sec:systematics}).  Systematic
errors are estimated separately for each detector configuration.  The energy distributions 
for events in the IC59 through IC86-IV data sets are similar and match the distribution 
shown in the right panel of Fig.\ref{fig:energy_dist}.

Within errors, the large-scale structure is stable over the data taking period 
considered here.  Table~\ref{tab:tdependence} shows the $\chi^2$ values calculated 
by comparing each year to the ensemble.  The resulting $p$-values are consistent with 
random fluctuations, indicating there is no time dependence over the period of this study.  
In addition, no systematic trends with time are detected within the individual right 
ascension bins in Fig.\,\ref{fig:proj1dcomp}.  A study of the stability over a period of 
twelve years (2000-2012) using data recorded with the AMANDA and IceCube 
detectors~\citep{Aartsen:2013lla} with the same method also did not find evidence
for a time dependence of the structure.

\begin{table}[ht]
  \centering
  \begin{tabular}{crrr}
    \hline
    Configuration & $\chi^2$ & $N$ & $p$-value\\
    \hline
    IC59     &     31.42 &     23 &      0.11 \\
    IC79     &     16.12 &     23 &      0.85 \\
    IC86     &     19.66 &     23 &      0.66 \\
    IC86-II  &     15.44 &     23 &      0.88 \\
    IC86-III &     24.27 &     23 &      0.39 \\
    IC86-IV  &     18.16 &     23 &      0.75 \\
    \hline
  \end{tabular}
  \caption{Results of a $\chi^2$ test comparing the relative intensity profile 
(see Fig.\,\ref{fig:proj1dcomp}) for each year to the collective ensemble.
The table provides $\chi^2$ values, number of degrees of freedom, $N$, 
and corresponding $p$-values.}
  \label{tab:tdependence}
\end{table}
    
To study the time dependence of the small-scale structure, we analyze the relative
intensity of the excess and deficit regions listed in Tab.\,\ref{tab:minmax} as a 
function of time.  For the location of the regions, we use the values determined from 
the full six-year data set.  Fig.\,\ref{fig:time_small} shows the relative intensity 
for each detector configuration, i.e., as a function of time, for each region.  
Also shown is the average value as determined from the analysis of the full six-year
data set.  The error bars on the data points and the error band on the average indicate 
statistical uncertainties only, but we list the average flux, including statistical
and systematic errors, for each region in the figures.  The systematic errors for the 
individual years have similar values.  The relative intensity at the excess and deficit
regions of the small-scale structure is constant within errors for the time period covered
by this analysis.  

As an additional test of the stability of the small-scale map, we subtract the relative 
intensity sky map of the full six-year data set from the sky map of each individual 
detector configuration, i.e., of each of the six years of data, and calculate the 
angular power spectrum of the residual maps.  All of them have power spectra that
are, within errors, compatible with isotropy, indicating that there are no significant
differences between the maps for individual years and the average.  The small-scale 
anisotropy, like the large-scale anisotropy, is constant over the time period covered 
by this analysis.

\begin{figure*}[ht]
  \centering
    \includegraphics[width=0.8\textwidth]{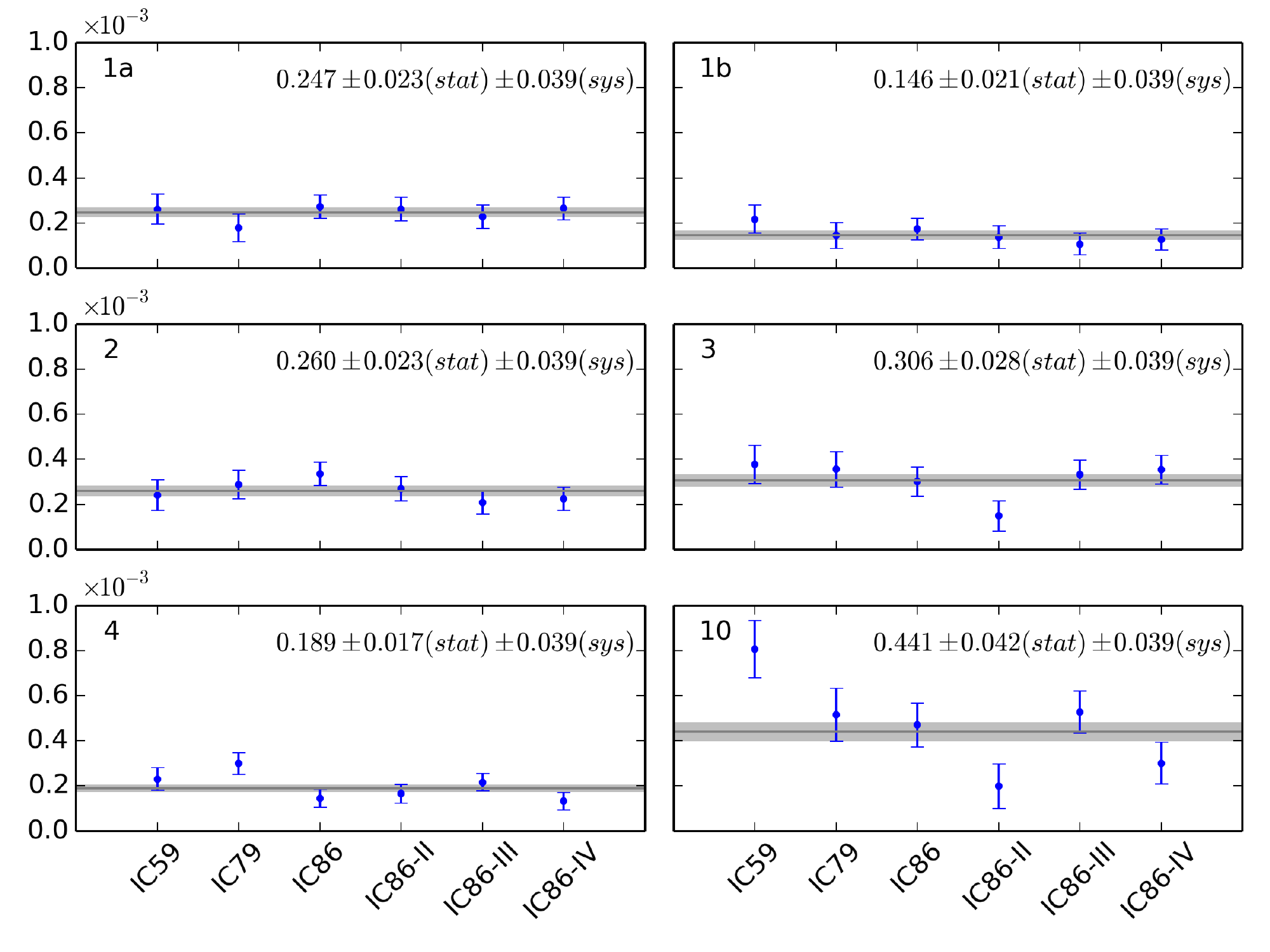}
    \includegraphics[width=0.818\textwidth]{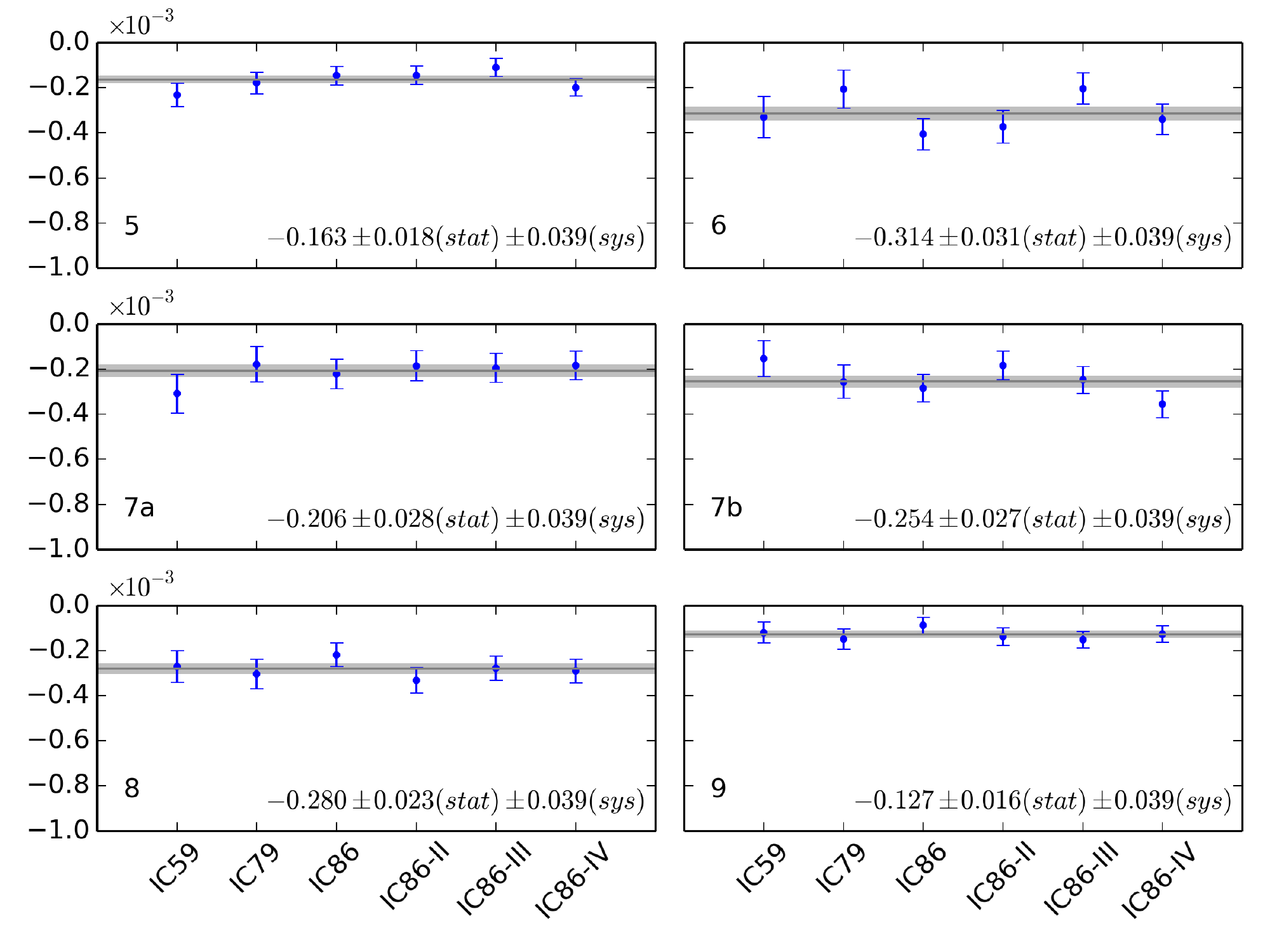}
    \caption{Relative intensity as a function of detector configuration for the locations
of the excess and deficit regions of the small-scale structure listed in Tab.\,\ref{tab:minmax}.
The horizontal lines indicate the six-year average relative intensity.  Error bars and bands 
are statistical, but the relative intensity, including statistical and systematic errors, 
is given for each region.}
    \label{fig:time_small}
\end{figure*}

\section{Systematic Checks}
\label{sec:systematics}

\begin{figure*}[ht]
  \centering
    \includegraphics[width=0.49\textwidth]{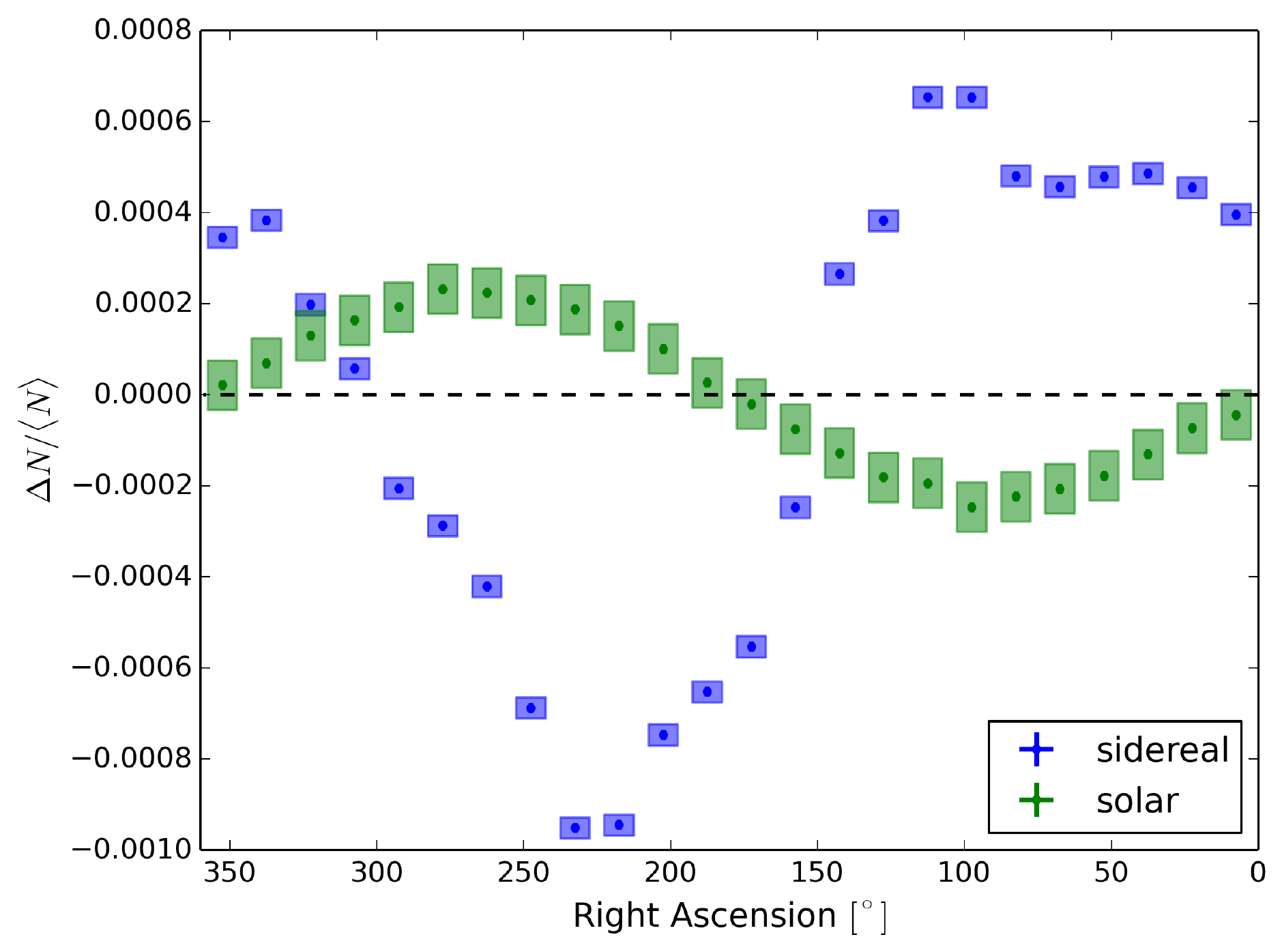}
    \includegraphics[width=0.49\textwidth]{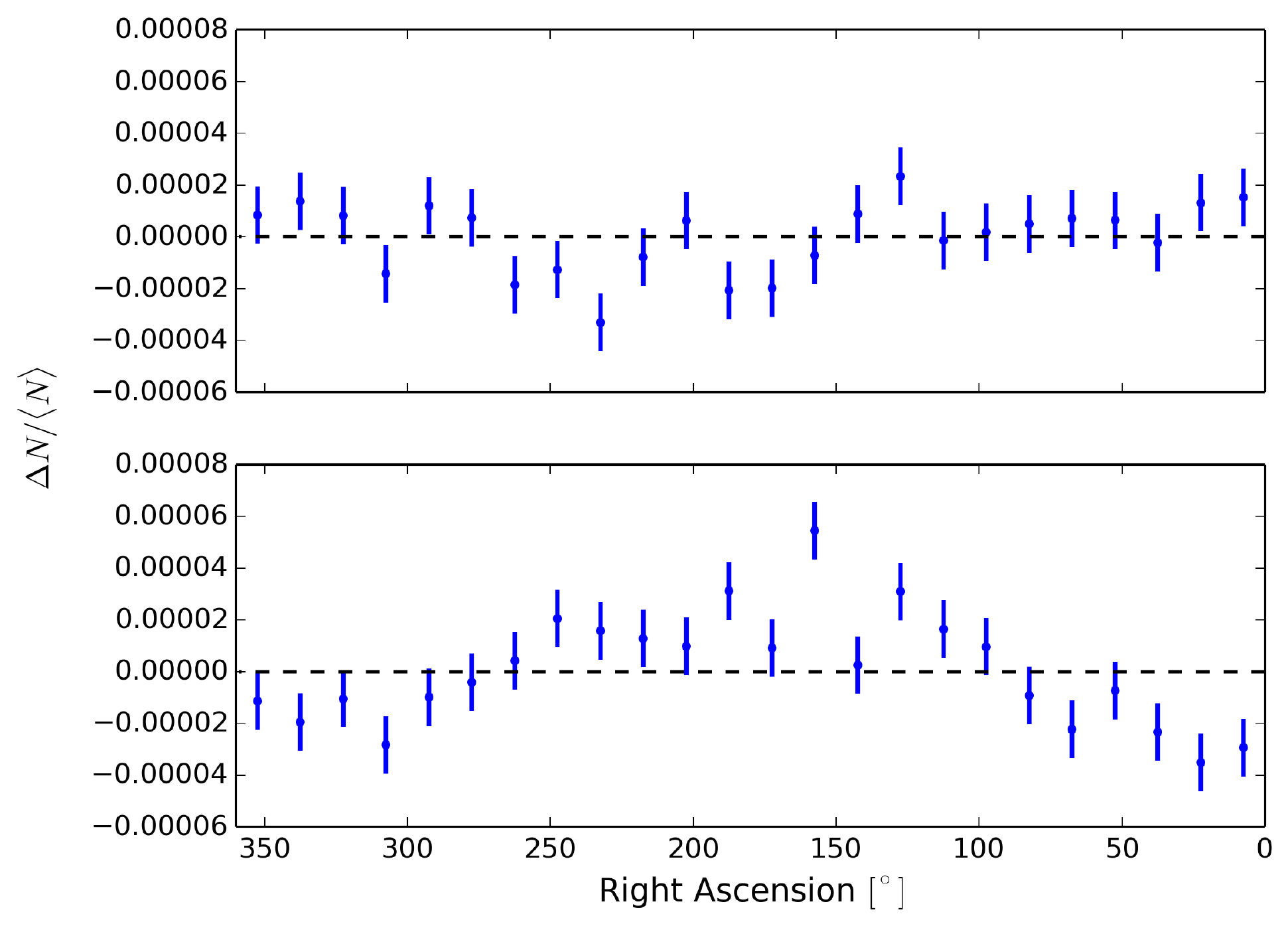}
    \caption{Projection of relative intensity in right ascension for
solar and sidereal time (\textit{left}), and anti-sidereal (\textit{right, top})
and extended-sidereal time (\textit{right, bottom}).  Error boxes for the solar 
and sidereal projection indicate systematic errors.  Note that both the anti-sidereal
and extended-sidereal frames are non-physical, so ``right ascension'' has no physical 
meaning in these frames.  For the solar frame, the right ascension axis shows the
difference between the right ascension of the event and the right ascension of the Sun.} 
    \label{fig:proj1d}
\end{figure*}

In~\citet{IceCube:2010aug}, several sources of systematic bias are considered,
including detector geometry and livetime, nonuniform exposure to different
regions of the sky, and seasonal variations in atmospheric conditions.  The
location of the IceCube detector minimizes the effect of some of these sources; 
the southern celestial sky is fully visible at all times, and seasonal 
variations are slow and automatically accounted for in the estimation of 
the reference map.  The checks performed in that previous analysis continue 
to hold, and the detector livetime has improved on average, as seen
in Tab.\,\ref{tab:data}.  In this section we expand on one possible source of
systematic bias that the increased data set allows us to study in more detail:
the possible influence of the solar dipole on the sidereal signal and
vice versa.

As the Earth orbits around the Sun, we observe an excess in the relative
intensity of cosmic rays in the direction of motion and a corresponding deficit
in the direction opposite to the motion.  This effect manifests itself as a
dipole in the relative intensity when the cosmic-ray arrival directions 
are plotted using solar time, i.e., in a frame where the position of the
Sun is at a fixed location.  This solar dipole has been measured 
previously~\citep{IceCube:2011oct, IceCube:2012feb} and now serves as a 
check of the consistency and reliability of the analysis methods used.

Ideally, the solar dipole should not cause any systematic uncertainties in the 
analysis of cosmic-ray arrival directions in sidereal time, as any signal in solar
time averages to zero over a year.  In practice, however, seasonal variations
in the solar dipole can manifest themselves as an anisotropy in the sidereal 
time frame and vice versa.  In order to study this mutual influence, we consider 
two nonphysical time scales: anti- and extended-sidereal time.  

Solar time has a frequency of 365.24 cycles per year.  The 
sidereal day is roughly four minutes shorter, with a frequency of 366.24 cycles 
per year.  The influence of the solar dipole on the sidereal anisotropy 
can be estimated from the influence it has on the other side band in frequency 
space, i.e., on a frame with 364.24 cycles per year.  This is the anti-sidereal
frame.  No physical signal is expected with a frequency of 364.24 cycles per year, 
so any significant ``signal'' that appears in the anti-sidereal frame stems from
a modulation of the solar frame and is equivalent to the systematic effect of the 
solar frame on the sidereal frame and therefore on the anisotropy signal.
This method can also be used to estimate the effect of the sidereal anisotropy
on the solar dipole.  In this case, the second side band of the sidereal frame,
the extended-sidereal time frame with a frequency of 367.24 cycles per year, can 
be analyzed.  Any significant signal in the extended-sidereal frame is due to 
seasonal modulations in the sidereal frame and is equivalent to the systematic 
effect of the sidereal frame on the solar frame.

The projection of the relative intensity in right ascension for the sidereal and 
solar frames is shown in Fig.\,\ref{fig:proj1d} (\textit{left}).  For the solar
frame, the ``right ascension'' axis shows the difference between the right ascension 
of the event and the right ascension of the Sun.  In this system, the Sun is located 
at $0^{\circ}$, so the maximum of relative intensity is at an angle of $270^{\circ}$, 
in the direction of the Earth's motion.  The solar dipole is well measured with six 
years of IceCube data.  The fit of the projection to a dipole results in an amplitude 
of $(2.242\pm 0.029)\times 10^{-4}$ and a phase of $(268.00\pm 0.75)^{\circ}$.  
The $\chi^2$-probability of the fit is 0.21 ($\chi^2=17.42$ for 23 degrees of 
freedom).  The amplitude of the sidereal anisotropy is larger, but is not well 
described by a dipole.  Statistical errors are shown, but are smaller than the 
data points due to the high statistics of the data set.  

In contrast, Fig.\,\ref{fig:proj1d} (\textit{right}) shows the projection of
the relative intensity in anti-sidereal and extended-sidereal time.  Note that
the amplitude of these projections is an order of magnitude smaller than the
amplitude of the solar dipole and the sidereal anisotropy, indicating that the 
effect of the solar on the sidereal frame and vice versa is small.  
We use the maximum amplitude of the relative intensity in the anti- and 
extended-sidereal frames as a conservative estimate for the systematic error in 
the sidereal and solar frame resulting from the other.  This 
amplitude appears as systematic error bars in Fig.\,\ref{fig:proj1dcomp}
and Fig.\,\ref{fig:proj1d} (\textit{left}).

\section{Summary and Discussion}
\label{sec:discussion}

\subsection{Large-Scale Anisotropy}

The analysis of 318 billion cosmic-ray events recorded between May 2009 and May 2015 
has shown anisotropy in the arrival direction distribution consistent with previously 
published IceCube results~\citep{IceCube:2011oct, IceCube:2012feb}.  The increased 
statistics of this data set allow for observation of the small-scale structure at a 
level approaching the angular resolution.  The resulting sky map shows separate 
structures that were not resolved in previous analyses, as well as two new regions, 
an excess and a deficit, observed with high statistical significance.

In addition, a detailed study of the evolution of the anisotropy as a function of
energy in the TeV to PeV range shows a strong dependence of the amplitude and the 
morphology of the anisotropy on energy.  This analysis extends our previous 
work~\citep{IceCube:2011oct, IceCube:2013mar} and confirms that the anisotropy 
changes rather dramatically between 130\,TeV and 240\,TeV; the phase of a best-fit 
dipole shifts from around $50^{\circ}$ to $270^{\circ}$ in right ascension.  At 
energies below this shift, the amplitude of the best-fit dipole decreases.  Above 
the shift, it increases again, up to the highest energies currently accessible to 
IceCube.

The source of the cosmic-ray anisotropy remains unknown.  The large-scale anisotropy 
may be qualitatively explained by homogeneous and isotropic diffusive 
propagation of cosmic rays in the Galaxy from stochastically distributed sources.  
Such discrete sources are responsible for a density gradient of cosmic rays, which 
causes a dipole anisotropy.  Numerical studies show that it is possible to find a 
particular realization of Galactic source distribution that explains the observed 
non-monotonic energy dependence of the anisotropy amplitude.  The change in the
phase of the anisotropy between TeV and PeV energies could indicate that the location 
of the dominant source(s) shifts from the Orion arm to the direction of the Galactic 
center~\citep{Sveshnikova:2013dec}.  The observed phase above several hundred TeV
coincides with the right ascension of the Galactic center, $\alpha_{GC}=268.4^{\circ}$.

As indicated earlier, the Pierre Auger Observatory has studied the amplitude and phase 
of the first harmonic modulation in right ascension at EeV energies~\citep{Auger:2011mar, 
Auger:2012dec, Auger:2013jan, Aab:2015bza}.  While the amplitude did not show any 
significant deviation from isotropy, the phase measurement showed consistent results 
in adjacent energy bins.  This was interpreted as a first indication of 
anisotropy.  At energies below 1\,EeV, a phase of $270^{\circ}$ for the first harmonic 
was found, a result that is consistent with the phase measured at PeV energies by IceCube 
and IceTop (see Fig.\,\ref{fig:dipole}).  Around 1\,EeV, a phase shift occurs, and above 
4\,EeV, the phase is about $100^{\circ}$.  Since this is roughly consistent with the 
direction towards the Galactic anticenter, the shift might be caused by a transition 
from a Galactic to an extragalactic origin of cosmic rays.  The gap between the IceCube
measurements and the measurements by the Pierre Auger Observatory is filled by the 
KASCADE-Grande experiment.  KASCADE-Grande data shows a dipole phase between median 
energies of 2.7\,PeV and 33\,PeV which is consistent with the IceCube results at PeV
energies and the Auger results below 1\,EeV~\citep{KASCADE:2015icrc}.  For the dipole 
amplitude, KASCADE-Grande measurements only yield upper limits.

While the interpretation of the dipole phase as an indication of the direction
towards the dominant source or sources is tantalizing, simulations of large-scale
phase and amplitude resulting from certain source distributions show that for an 
ensemble of realizations, the mean amplitude is larger than what is 
observed~\citep{Erlykin:2006apr, Blasi:2012jan, Ptuskin:2012dec, Pohl:2013mar, 
Sveshnikova:2013dec}.  It is known that transport of cosmic rays in magnetic fields 
is anisotropic, even for large magnetic perturbations compared to the regular mean 
field~\citep{Giacalone:1999, Shalchi:2009, Tautz:2009, Desiati:2014jya, Shalchi:2015} 
and that propagation perpendicular to the local magnetic field direction is slower 
than in the parallel direction.  The possible misalignment between the regular magnetic 
field and the cosmic-ray density gradient decreases the amplitude of the observed 
anisotropy.  This might explain the observed smaller amplitudes of the 
anisotropy~\citep{Effenberger:2012nov, Kumar:2014apr, Mertsch:2015jan}, but it also 
means that the anisotropy does not point in the direction of any particular nearby 
source.

The fact that the cosmic-ray anisotropy is not a simple dipole, nor well fit solely
by lower-multipole terms, suggests that other transport processes might be
important as well.  For instance, drift diffusion driven by a gradient of
cosmic-ray density in the local interstellar medium, producing a bidirectional
flow of Galactic cosmic rays, was considered by~\citet{Tibet:2007aug} 
and~\citet{Mizoguchi:2009sep}.

\subsection{Small-Scale Anisotropy}

The small-scale anisotropy may be produced by the interactions of cosmic rays
with an isotropically turbulent interstellar magnetic field.  Scattering
processes with stochastic magnetic instabilities produce perturbations in the
arrival direction distribution of an anisotropic distribution of cosmic-ray
particles within the scattering mean free path.  Such perturbations may be
observed as stochastic localized excess or deficit regions~\citep{Giacinti:2012aug, 
Biermann:2013may}.  The corresponding angular power spectrum can be analytically 
predicted from Liouville's theorem~\citep{Ahlers:2014jan, Ahlers:2015dwa}.  The 
injection scale of interstellar turbulence is on the order of 10\,pc within the 
Galactic arms and 100\,pc in the inter-arm regions~\citep{Haverkorn:2006jan}.  
In the cascading processes down to smaller scales, the turbulent eddies become 
elongated along the magnetic field lines.  This anisotropic turbulence makes 
scattering processes inefficient.  The scattering mean free path can be larger 
than the turbulence injection scale, so particles basically stream along magnetic 
field lines with small cross-field line transport~\citep{Yan:2008feb, Lazarian:2014mar}.

Besides the cascading interstellar magnetic field turbulence down to the damping
scale (typically on the order of 0.1\,pc), there are other sources of magnetic
perturbations on smaller scales.  The closest to Earth is represented by the
heliosphere, formed by the interaction between the solar wind and the
interstellar flow.  It is about 600 astronomical units (AU) wide, and it could
extend several thousand AU downstream of the interstellar
wind~\citep{Pogorelov:2009may}.  Globally, the heliosphere constitutes a
perturbation in the 3\,$\mu$G local interstellar magnetic field with an
injection scale comparable to the $\sim$\,10\,TeV proton gyroradius (0.003\,pc, or
620 AU).  It is therefore reasonable that the local interstellar 
magnetic field draping around the heliosphere might be a significant source of 
resonant scattering, capable of redistributing the arrival directions of TeV 
cosmic-ray particles.

\subsection{Time Dependence}

A study of the time dependence of the large- and small-scale structure
over the six-year period covered by this analysis reveals no significant change with 
time.  This result is consistent with previous studies in the Northern and Southern 
Hemispheres~\citep{ARGO:2013oct, Tibet:2010mar, Aartsen:2013lla}, but inconsistent
with others~\citep{Nagashima:1998aug,Milagro:2008nov}.

We do not expect time variations due to interstellar magnetic field effects,
so the non-observation of time variations does not constrain any astrophysical
scenarios that could explain the cosmic-ray anisotropy.  However, time variations
are expected to arise from local phenomena within the heliosphere.  At TeV energies,
we are not sensitive to the effects of the solar wind and to the direct effects
of the solar cycles.  Such direct effects, caused by the modulation of the solar
wind average strength and its interaction with the cosmic rays, are only detectable
at energies below 30\,GeV.  Modulations in the cosmic-ray spectrum below 30\,GeV
arising from the 11-year solar cycle have been observed by other 
experiments~\citep{Potgieter:2014,Potgieter:2015,Adriani:2015}.

On the other hand, there could be an indirect effect of the solar cycles on
higher energy cosmic rays.  This might arise from the fact that the reversing
in the solar magnetic field polarity every 11 years produces vast regions of
magnetized plasma pushed away from the Sun straight into the heliospheric tail
by the solar wind.  In such regions the magnetic field is expected to have a
coherent polarity~\citep{Pogorelov:2009may}.  Subsequent 11-year solar cycles, 
therefore, produce a series of reversing uni-polar magnetic field regions that 
are pushed towards the heliotail.  Each region is approximately 200 to 300\,AU 
wide, roughly the proton gyroradius at TeV energies in typical Galactic magnetic 
fields.  Cosmic rays, especially with high $Z$, might be affected by these 
magnetized regions, in particular those produced in the previous solar cycle 
and just past the solar system towards the heliotail~\citep{Lazarian:2010oct,
Desiati:2012jun,Desiati:2013jan}.  We should observe an 11-year modulation, 
although not necessarily in sync with the current solar cycle.  However, based
on current observations, such effects may be very small, possibly smaller than 
$10^{-5}$ in relative intensity, and they might only be detectable towards the 
general direction of the heliotail.

Annual modulations may also be expected from the fact that in December, Earth is
slightly closer to the heliotail than in June.  However, the variability in
relative intensity is also expected to be of the order of $10^{-5}$ or less 
and cannot be detected or excluded with IceCube data.

\subsection{Outlook}

The PeV energy region is not only significant for the energy-dependent cosmic-ray
anisotropy; it is also a region where the cosmic-ray energy spectrum shows noticeable
fine structure and the chemical composition of the cosmic-ray flux changes (see
for example~\citet{Aartsen:2013wda}).  In the future, we will focus on a detailed 
study of possible connections between the arrival direction anisotropy and the 
energy spectrum and chemical composition of the cosmic-ray flux.  The IceTop air 
shower array has an energy resolution better than 0.1 in 
$\log_{10}(E/\mathrm{GeV})$~\citep{IceCube:2013feb}.  IceTop data can thus be 
used to compare the energy spectrum in regions of excess or deficit flux to 
the isotropic spectrum.  At TeV energies, this type of analysis already showed 
that the spectrum in excess regions is harder than the overall energy 
spectrum~\citep{Milagro:2008nov, ARGO:2013oct, HAWC:2014dec}.  
With IceTop data, we can search for similar effects at PeV energies.  
IceTop also has some sensitivity to the chemical composition of the cosmic-ray flux.  
A study of composition-dependent parameters as a function of sky location could 
reveal correlations between the anisotropy and the composition of the cosmic-ray 
flux.  In addition, future data will help to extend the IceCube/IceTop measurements 
to higher energies where they can be compared with results from the KASCADE-Grande
experiment in the Northern Hemisphere.

With new cosmic-ray data of unprecedented quantity and quality now available from a 
number of experiments, the challenge for any theory of cosmic-ray origin and propagation 
is to explain simultaneously the fine structure of the cosmic-ray energy spectrum, 
the chemical composition of the cosmic-ray flux, and the amplitude and phase of the 
anisotropy over a wide energy range from TeV to EeV energies.

\acknowledgments

We acknowledge the support from the following agencies:
U.S. National Science Foundation-Office of Polar Programs,
U.S. National Science Foundation-Physics Division,
University of Wisconsin Alumni Research Foundation,
the Grid Laboratory Of Wisconsin (GLOW) grid infrastructure at the University of Wisconsin - Madison, 
the Open Science Grid (OSG) grid infrastructure;
U.S. Department of Energy, and National Energy Research Scientific Computing Center,
the Louisiana Optical Network Initiative (LONI) grid computing resources;
Natural Sciences and Engineering Research Council of Canada,
WestGrid and Compute/Calcul Canada;
Swedish Research Council,
Swedish Polar Research Secretariat,
Swedish National Infrastructure for Computing (SNIC),
and Knut and Alice Wallenberg Foundation, Sweden;
German Ministry for Education and Research (BMBF),
Deutsche Forschungsgemeinschaft (DFG),
Helmholtz Alliance for Astroparticle Physics (HAP),
Research Department of Plasmas with Complex Interactions (Bochum), Germany;
Fund for Scientific Research (FNRS-FWO),
FWO Odysseus programme,
Flanders Institute to encourage scientific and technological research in industry (IWT),
Belgian Federal Science Policy Office (Belspo);
University of Oxford, United Kingdom;
Marsden Fund, New Zealand;
Australian Research Council;
Japan Society for Promotion of Science (JSPS);
the Swiss National Science Foundation (SNSF), Switzerland;
National Research Foundation of Korea (NRF);
Villum Fonden, Danish National Research Foundation (DNRF), Denmark

\end{document}